\documentclass[a4paper,11pt]{article}
\pdfoutput=1 
\usepackage{jheppub} 
\usepackage[T1]{fontenc} 
\usepackage{graphicx}
\graphicspath{{fig/}} \DeclareGraphicsExtensions{.pdf}
\usepackage{amsmath}
\usepackage{bm}
\usepackage{slashed}
\usepackage{amsfonts}

\def\s0{\sigma_0}
\def\beq{\begin{equation}}
\def\eeq{\end{equation}}
\def\bear{\begin{eqnarray}}
\def\enar{\end{eqnarray}}

\newcommand{\state}[4]{{^#1\hspace{-0.6mm}#2_{#3}^{[#4]}}}

\newcommand\lamTheta{ \lambda_{\theta} }
\newcommand\lamPhi{\lambda_{\phi}}
\newcommand\lamThPh{\lambda_{\theta\phi}}
\newcommand\lamINV{\lambda_{\rm{inv}}}

\newcommand\gev{\mathrm{~GeV}}
\newcommand\tev{\mathrm{~TeV}}

\newcommand{\jpsi}{{J/\psi}}

\newcommand\CScSa{\state{3}{S}{1}{1}}

\newcommand\COaSz{\state{1}{S}{0}{8}}

\newcommand\COcSa{\state{3}{S}{1}{8}}
\newcommand\COcPz{\state{3}{P}{0}{8}}

\newcommand\COcPj{\state{3}{P}{J}{8}}

\newcommand\mo{{\mathcal O}}

\newcommand{\LDME}[2]{\langle\mo^{#1}(#2)\rangle}
\newcommand\mops{\LDME{\jpsi}{\CScSa}}
\newcommand\mopa{\LDME{\jpsi}{\COaSz}}
\newcommand\mopb{\LDME{\jpsi}{\COcSa}}
\newcommand\mopc{\LDME{\jpsi}{\COcPz}}

\newcommand{\vt}[1]{{{\boldsymbol #1}_\perp}}
\newcommand{\vtn}[2]{{{\boldsymbol #1}_{#2\perp}}}

\newcommand{\vx}{{\vt{x}}}
\newcommand{\vy}{{\vt{y}}}

\newcommand{\vp}{{\vt{p}}}

\newcommand{\vk}{{\vt{k}}}
\newcommand{\vka}{{\vtn{k}{1}}}

\newcommand{\vtp}[1]{{{\boldsymbol #1}'_\perp}}

\newcommand{\vkp}{{\vtp{k}}}

\title{\boldmath $J/\psi$ polarization in the CGC+NRQCD approach}

\author[a,b,c]{Yan-Qing Ma,}
\author[d,e]{Tomasz Stebel,}
\author[e]{and Raju Venugopalan}

\affiliation[a]{School of Physics and State Key Laboratory of Nuclear Physics and
Technology, Peking University, Beijing 100871, China.}
\affiliation[b]{Center for High Energy Physics,
Peking University, Beijing 100871, China.}
\affiliation[c]{Collaborative Innovation Center of Quantum Matter,
Beijing 100871, China.}

\affiliation[d]{Institute of Nuclear Physics PAN, Radzikowskiego 152, 31-342 Krak\'ow, Poland.}

\affiliation[e]{Physics Department,
Brookhaven National Laboratory,
Upton, NY 11973, USA.}

\emailAdd{yqma@pku.edu.cn}
\emailAdd{tomasz.stebel@ifj.edu.pl}
\emailAdd{raju.venugopalan@gmail.com}

\date{\today}

\abstract{
We compute the $J/\psi$ polarization observables $\lamTheta$, $\lamPhi$, $\lamThPh$ in a Color Glass Condensate (CGC) + nonrelativistic QCQ (NRQCD) formalism that includes contributions from both color singlet and color octet intermediate states. Our results are compared to low $p_T$ data on $J/\psi$ polarization from the LHCb and ALICE experiments on proton-proton collisions at center-of-mass energies of $\sqrt{s}=7\tev$ and 8 TeV. 
Our CGC+NRQCD computation provides a better description of data for $p_T \leq 15$ GeV relative to extant next-to-leading (NLO) calculations within the collinear factorization framework. These results suggest that higher order computations in the CGC+NRQCD framework have the potential to greatly improve the 
accuracy of extracted values of the NRQCD universal long distance matrix elements. 
}

\begin{document}
\maketitle

\section{Introduction}

The study of heavy quarkonium states in QCD is an essential ingredient in developing our understanding of the subtle interplay of short and long distance physics in QCD. However even though the simplest $J/\psi$ meson was discovered more than 40 years ago, key features of how this state is produced in high energy collisions continue to elude us. A prominent example is the polarization of the $J/\psi$, which appeared to differ significantly from theoretical expectations. 

The theoretical models describing the production of heavy flavors in QCD rely on the factorization between the hard process governing the production of the 
heavy quark-antiquark pair and the soft processes governing the hadronization of this pair into quarkonium states such as the $J/\psi$. The former can be computed in perturbative QCD while the latter is intrinsically nonperturbative and can be determined only from models or effective field theory approaches. For instance, in the color evaporation model \cite{Fritzsch:1977ay}, the perturbative production of the heavy quark-antiquark ($Q\bar Q$) pair with mass $M$ is followed by its nonperturbative hadronization to the final state meson with a universal transition probability for all $M$ below the mass threshold of producing two open flavor heavy mesons. In the color singlet model, the $Q\bar Q$ pair is produced in a color singlet state before hadronizing into the quarkonium state. The $Q\bar Q$ wave function in this approach is computed at zero separation between the quark-antiquark pair. The most sophisticated approach to describe the hadronization of heavy quarkonia is nonrelativistic QCD~\cite{Bodwin:1994jh}, an effective field theory valid in the limit of very heavy quark masses. NRQCD employs systematic power counting in the relative velocity of the $Q\bar Q$ to determine the long distance matrix elements (LDMEs) of the dominant nonpeturbative operators contributing to the formation of quarkonia. For the $J/\psi$, the LDME of the color singlet channel is dominant in the power counting of the matrix elements, with significant contributions also arising from several color octet channels \cite{Bodwin:1994jh,Cho:1995vh,Cho:1995ce}. 

There has been a large amount of work in recent years developing the NRQCD formalism and applying it to quarkonium measurements in a wide range of experiments~\cite{Brambilla:2010cs}. Our focus will be on quarkonium production in proton-proton collisions. In this case, the matrix elements for the production of $Q\bar Q$ pairs in both color singlet and color octet states were calculated within the collinear factorization formalism including next-to-leading (NLO) perturbative corrections. These results were employed by several groups to extract the nonperturbative LDMEs from comparisons of the cross-sections to data~\cite{Gong:2008ft,Ma:2010yw,Ma:2010jj,Butenschoen:2010rq,Butenschoen:2011yh}. 

For the production of $J/\psi$'s with large transverse momentum $p_T$, the most important contribution in NRQCD at leading order (LO) in $\alpha_s$ comes from the $\COcSa$ channel, which suggests that the produced $J/\psi$'s should be transversely polarized \cite{Braaten:1999qk}. 
The polarization of the $J/\psi$ is extracted from the angular distribution of positively charged (by convention) leptons in the decay of $J/\psi$ into muons ($J/\psi\rightarrow \mu^+ \mu^-$), that is parametrized by the coefficients $\lamTheta$, $\lamPhi$ and $\lamThPh$. Transversally (longitudinally) polarized $J/\psi$'s have $\lamTheta=1 (-1)$, $\lamPhi,\lamThPh=0$ while all coefficients are zero for unpolarized $J/\psi$'s. Measurements of the $J/\psi$ polarization
by the CDF Collaboration at the Tevatron \cite{Affolder:2000nn,Abulencia:2007us}, as well as the ALICE \cite{Abelev:2011md,Acharya:2018uww}, LHCb \cite{Aaij:2013nlm} and CMS \cite{Chatrchyan:2013cla} experiments at the LHC, showed that the $J/\psi$ has weak or no polarization. This stark disagreement between the leading NRQCD expectation and collider data has been dubbed the "$J/\psi$ polarization puzzle". 

Extensions of the LO NRQCD $J/\psi$ polarization studies in hadronic collisions to next-to-leading order (NLO) in collinearly factorized perturbative QCD (pQCD) approaches have been discussed in \cite{Butenschoen:2012px,Chao:2012iv}; further studies including feeddown contributions from higher states were also discussed in \cite{Gong:2012ug,Shao:2014yta}. The latter computations are important because there are no available experimental data for direct $J/\psi$ production; the polarization is measured either for "prompt" (including feeddown from the higher excited charmonium states $\psi(2s), \ \chi_{cJ} \ldots$) \cite{Affolder:2000nn,Abulencia:2007us,Aaij:2013nlm,Chatrchyan:2013cla} or "inclusive" production (prompt plus additional contributions from bottom meson decays) \cite{Abelev:2011md,Acharya:2018uww}. The conclusion of these NLO computations was that the $\COcSa$ channel and $\COcPj$ channels have a large cancellation between their transverse and longitudinal polarization components \cite{Chao:2012iv}, thereby providing a possible explanation for the lack of $J/\psi$ polarization at high $p_T$.

However the $J/\psi$ produced in proton-proton collisions are also weakly polarized at low values of $p_T$ where the collinear factorization formalism may not be applicable. At low $p_T$'s at collider energies, large $\alpha_s \ln(1/x)$ contributions arise at higher orders that may not be fully accounted for in collinear factorization frameworks. Another source of $O(1)$ contributions are higher twist multiparton matrix elements that are large at low $p_T$. Small $x$ kinematics is also accessed in either the projectile or target at forward rapidities. For instance, the LHCb \cite{Aaij:2013nlm} and ALICE \cite{Abelev:2011md,Acharya:2018uww} experiments measure $J/\psi$'s at the forward rapidities of $2<y<4.5$ (LHCb) and $2.5<y<4$ (ALICE) for $p_T < 15\gev$, providing access to $x$ values down to $x\sim 10^{-4}$ in one of the protons. 

The contribution of large $\alpha_s \ln(1/x)$ contributions as well as leading higher twist contributions to quarkonium production can be computed systematically in the Color Glass Condensate (CGC) effective field theory (EFT) \cite{Iancu:2003xm,Weigert:2005us,Gelis:2010nm,Kovchegov:2012mbw,Blaizot:2016qgz}. This EFT treats large $x$ degrees of freedom in the two hadrons as static color sources that are coupled to dynamical gauge fields 
at small $x$. Physical quantities such as heavy quark pair cross-sections are computed in a two-step procedure;  they are first computed for a fixed distribution of color sources, in the gauge field background, and subsequently averaged over a gauge invariant distribution of color sources. The separation scale in $x$ between large $x$ sources and small $x$ fields is arbitrary. However the requirement that physical quantities do not depend on this separation scale leads to a renormalization group equation, the JIMWLK equation~\cite{JalilianMarian:1997gr,JalilianMarian:1997dw,Iancu:2000hn,Ferreiro:2001qy}, describing the change in the distribution of color sources with decreasing $x$. A key feature of this approach is a dynamically generated saturation scale $Q_s(x)$ \cite{Gribov:1984tu,Mueller:1985wy,McLerran:1993ni,McLerran:1993ka} that grows both with decreasing $x$ and with increasing nuclear size. When $Q_s^2 \gg \Lambda_{\rm QCD}^2$, one can employ weak coupling methods to compute cross-sections even at low $p_T$ since $\alpha_S(Q_s^2)\ll 1$. 

The heavy quark pair production cross-section for proton-proton and proton-nucleus collisions was computed in a "dilute-dense" approximation of the CGC in \cite{Blaizot:2004wv,Fujii:2006ab}. This approximation corresponds to systematically keeping lowest order terms in an expansion of the smaller of the color charge densities of the two colliding hadrons, and terms to all orders in the larger of the two color charge densities -- hence the moniker dilute-dense. It is strictly valid for forward proton-proton collisions or in proton-nucleus collisions in $p_T$ and rapidity windows that are consistent with this expansion\footnote{This can only be estimated a priori; strictly speaking, only a computation of the next-order correction can assess the accuracy of the approximation.}. The dilute-dense results for heavy quark pair production were later used to compute the short distance cross-section (SDC) for Onium production in a CGC+NRQCD approach \cite{Kang:2013hta}. 
Numerical results for $J/\psi$ production in $p+p$ at RHIC and LHC collisions were presented in \cite{Ma:2014mri} and likewise for $p+A$ in \cite{Ma:2015sia}. More recently, results for $J/\psi$ production in high multiplicity $p+p$ and $p+A$ collisions were obtained in \cite{Ma:2018bax}. The CGC+NRQCD approach describes quite well the systematics of the $p_T$ and rapidity dependence of the $J/\psi$ yields in both proton-proton and proton-nucleus collisions within theory uncertainties. 
We note that the CGC has been applied to compute the SDC in quarkonium production in hadron-hadron collisions\footnote{For other approaches to quarkonia production in $p+A$ collisions, see \cite{Albacete:2013ei,Vogt:2010aa,McGlinchey:2012bp,Arleo:2013zua}.} approximating the LDME with the color evaporation model~\cite{Fujii:2013gxa,Ducloue:2015gfa} and variants thereof~\cite{Ma:2016exq,Ma:2017rsu}.

In this paper, we shall extend the CGC+NRQCD analysis of \cite{Kang:2013hta} and \cite{Ma:2014mri} to address the $J/\psi$ polarization puzzle. 
We will begin by first relating the coefficients of the angular distribution of the positively charged leptons produced in $J/\psi$ leptonic decays to helicity dependent quarkonium cross-sections. These coefficients are frame dependent and are typically presented as such by the experiments though frame independent combinations of these can also be extracted. In Section \ref{den_matrix_elm_sect}, we will write down the explicit expressions for the helicity dependent SDCs in the CGC+NRQCD framework. Numerical results for the coefficients of the angular distribution are presented in Section \ref{num_res_sect}. We observe that the agreement of the theoretical computations with the data is 
quite good for the low $p_T$ LHC data. In particular, we observe that there is also a cancellation between the polarizations of the $\COcSa$ channel with that of the $\COcPj$ channel in the CGC+NRQCD framework, even for the LO in $\alpha_s$ impact factor. We will end with a summary and outlook on further work. Several details of the computation, and additional results, are provided in two appendices.

\section{Angular distribution of $J/\psi$ leptonic decay and helicity dependent cross-sections}
\label{ang_dist_section}

In order to extract the polarization of the $J/\psi$, first consider the leptonic decay of $J/\psi$ in its rest frame \cite{Lam:1978pu}. The angular $(\theta,\phi)$ distribution of the positive lepton is obtained by fixing a frame $X,Y,Z$ with respect to which the angles $\theta$ and $\phi$ are measured:
\bear
\label{l+_angles_definition}
p_{l^+}\cdot X&=& -\left|\vec p_{l^+}\right| \sin\theta\cos\phi \nonumber\\
p_{l^+}\cdot Y&=&-\left|\vec p_{l^+}\right| \sin\theta\sin\phi \\
p_{l^+}\cdot Z&=& -\left|\vec p_{l^+}\right| \cos\theta \,.\nonumber
\enar
Here $p_{l^+}$ is the four-momentum of the positively charged lepton created in the $J/\psi$'s decay and $\left|\vec p_{l^+}\right|$ is a length of its three-momentum in the $J/\psi$'s rest frame. The unit four-vectors $X,Y,Z$ span the subspace perpendicular to $J/\psi$'s momentum $p^\mu$ and are perpendicular to each other: $X\cdot p=0$, $Y\cdot p=0$, $Z\cdot p=0$, $X\cdot Y=0$, $X\cdot Z=0$, $Y\cdot Z=0$, $X^2= -1$, $Y^2= -1$, $Z^2= -1$. 

The orientation of the vectors $X, Y, Z$ with respect to the momenta $P_1$, $P_2$ of the incoming hadrons depends on a choice of frame. In this paper, we will consider two frames which are often used in the literature: the Collins--Soper \cite{Collins:1977iv} frame and the recoil (or helicity) frame \cite{Jacob:1959at}. In both frames, the $Y$ four-vector is chosen to be perpendicular to the hadron plane,
\begin{equation}
Y_\mu \propto \epsilon_{\mu\alpha\beta\gamma}
p^\alpha P_1^\beta P_2^\gamma\,. 
\label{Y_vector_def}
\end{equation}
In the $J/\psi$ center-of-mass frame, the corresponding three-vector is chosen to be\footnote{There is an additional freedom in choosing this vector to be aligned or anti-aligned with the 
positive Y-direction that needs to be specified by the experiment.},

\beq
\vec Y = \frac{-\vec P_1 \times \vec P_2}{\left|\vec P_1 \times \vec P_2\right|}\,.
\label{Y_orientation}
\eeq

In order to define the $X$ and $Z$ four-vectors, we first introduce projections of the hadron momenta on to the $J/\psi$'s four-momentum $p^\mu$:
\begin{eqnarray}
A=P_1+P_2 \qquad \tilde{A}^\mu=A^\mu-\frac{A\cdot p}{M^2} \, p^\mu,\nonumber
\\
B=P_1-P_2 \qquad \tilde{B}^\mu=B^\mu-\frac{B\cdot p}{M^2} \, p^\mu\,,
\label{A_B_vec_def}
\end{eqnarray}
where $p_\mu p^\mu = M^2$. Then $X$ and $Z$ are linear combinations of $\tilde{A}$ and $\tilde{B}$, 
\begin{eqnarray}
X^\mu&=&\alpha_x\tilde{A}^\mu+\beta_x\tilde{B}^\mu, \nonumber
\\
Z^\mu&=&\alpha_z\tilde{A}^\mu+\beta_z\tilde{B}^\mu.
\label{X_Z_def_by_A_B}
\end{eqnarray}
where the recoil and Collins--Soper frames are defined by the values of the coefficients $\alpha_{x,z}$, $\beta_{x,z}.$ Explicit expressions for these coefficients were given in \cite{Beneke:1998re}. For completeness, they are listed in Appendix \ref{appendix_alpha_beta_coeff}.

Since the $X$ and $Z$ vectors lie in the plane of the incoming hadrons, the positively charged lepton's angular distribution in the $J/\psi$ rest frame can be parameterized by three coefficients as \cite{Noman:1978eh} 
\beq
\label{ang_distribution_in_Jpsi_frame}
\frac{d\sigma^{J/\psi(\rightarrow l^+l^- )}}{d\Omega} \propto 1+\lamTheta\cos^2{\theta}+\lamPhi \sin^2\theta\cos2\phi+\lamThPh \sin 2\theta \cos\phi,
\eeq
where $\Omega=(\theta,\phi)$ denotes the solid angle of the positive lepton in Eq.~(\ref{l+_angles_definition}). The coefficients of this angular distribution computed using the spin density matrix elements for $J/\psi$ production can be expressed as~\cite{Noman:1978eh}  
\beq
\lamTheta=\frac{d\sigma_{11}-d\sigma_{00}}{d\sigma_{11}+d\sigma_{00}}\,, \hspace{1cm}
\lamPhi=\frac{d\sigma_{1,-1}}{d\sigma_{11}+d\sigma_{00}}\,, \hspace{1cm}
\lamThPh=\frac{\sqrt{2}\; \rm{Re}(d\sigma_{10}) }{d\sigma_{11}+d\sigma_{00}}\,.
\label{lam_Defin}
\eeq
The cross-section $d\sigma_{ij}$ corresponds to the product of the amplitude for inclusive production of a $J/\psi$ with helicity $i$ in the amplitude and helicity $j$ in the complex conjugate amplitude. Hence $d\sigma_{11}$ ($d\sigma_{00}$) can be interpreted as the cross-section for the production of $J/\psi$ with helicity $h=+1$ ($h=0$). The unpolarized cross-section is given by a sum of contribution of three helicity states: $h=+1$, $h=0$ and $h=-1$:
\beq
\label{unpol_xsect}
d\sigma=d\sigma_{11}+d\sigma_{00}+d\sigma_{-1-1}=2d\sigma_{11}+d\sigma_{00}\,,
\eeq
which is the trace of the spin density matrix.

The values of the coefficients $\lamTheta$, $\lamPhi$ and $\lamThPh$ depend on the choice of frame-- the choice of the $X$ and $Z$ axes. One can construct out of these coefficients frame-independent quantities as well, as discussed in \cite{Faccioli:2010ej,Faccioli:2010ji,Faccioli:2010kd,Palestini:2010xu,Ma:2017hfg}. We will present results for two of these invariants, which are defined as 
\beq
\lamINV^{(1)}= \frac{\lamTheta+3\lamPhi}{1-\lamPhi}, \hspace{1cm}
\lamINV^{(2)}= \frac{1+(\lamTheta-\lamPhi)/4}{\sqrt{(\lamTheta-\lamPhi)^2+4\lamThPh^2}}\,.
\label{lam_inv_definition}
\eeq

\section{Computation of the spin density matrix in CGC+NRQCD}
\label{den_matrix_elm_sect}

In what follows, we will write down the spin density matrix elements $d\sigma_{ij}$ in the CGC+NRQCD formalism. The spin density matrix elements can be expressed as~\cite{Bodwin:1994jh}
\beq
d\sigma_{ij}=\sum_\kappa d\hat\sigma^\kappa_{ij} \, \langle\mathcal O_\kappa\rangle\,,
\label{NRQCD_expansion}
\eeq
where $\langle\mathcal O_\kappa\rangle$ are the NRQCD long distance matrix elements (LDMEs). The SDCs $d\hat\sigma^\kappa_{ij}$ describe the production of the $c\bar c$ pair in a given quantum state $\kappa=\state{{2S+1}}{L}{J}{C}$, where $[C]$ denotes either the singlet $[1]$ or the octet $[8]$ color state.
The LDMEs describe the nonperturbative transition of the $c\bar c$ pair into the $J/\psi$ state; these are process independent and can be determined by fitting experimental data.

For $J/\psi$ production, the leading contribution to the sum in Eq.~(\ref{NRQCD_expansion}) comes from the states 
\beq
^3S_1^{[1]}, \ ^1S_0^{[8]}, \ ^3S_1^{[8]}, \ ^3P_J^{[8]} \textrm{ with } J=0,1,2\,.
\label{quantum_states_for_J/psi}
\eeq
Based on NRQCD velocity scaling rules \cite{Bodwin:1994jh}, the spin of the $J/\psi$ is the same as that of the intermediate $c\bar c$ pair if it is produced via $^3S_1^{[1]}$, $ ^3S_1^{[8]}$, or $^3P_J^{[8]}$ states. See also \cite{Tang:1995zp,Beneke:1998re} for further discussion.

In the CGC effective field theory, the SDC's are given by the expressions \cite{Kang:2013hta,Ma:2014mri},
\bear\label{eq:dsktCO}
\frac{d \hat{\sigma}_{ij}^\kappa}{d^2\vp d
y}&\overset{\text{CO}}=&\frac{\alpha_s (\pi R_p^2)}{(2\pi)^{7}
(N_c^2-1)} \underset{\vka,\vk}{\int}
\frac{\varphi_{p}(x_1,\vka)}{ k_{1\perp}^2} \mathcal{N}_Y(x_2,\vk) \nonumber \\
&\times&\mathcal{N}_Y(x_2,\vp-\vka-\vk)
\,\Gamma^{\kappa}_{ij}\left(x_1,x_2, p,\vka,\vk\right),
\enar
for the color octet channels and
\bear\label{eq:dsktCS}
\frac{d \hat{\sigma}_{ij}^\kappa}{d^2\vp dy}&\overset{\text{CS}} =&\frac{\alpha_s (\pi R_p^2)}{(2\pi)^{9}
(N_c^2-1)} \underset{\vka,\vk,\vkp}{\int}
\frac{\varphi_{p}(x_1,\vka)}{k_{1\perp}^2}\mathcal{N}_{Y}(x_2,\vk)\mathcal{N}_{Y}(x_2,\vkp) \\
&\times& \mathcal{N}_{Y}(x_2,\vp-\vka-\vk-\vkp)\,
{\cal G}^\kappa_{ij}\left(x_1,x_2, p,\vka,\vk,\vkp \right),
\enar
for the color singlet channels. In these expressions, $\mathcal{N}_{Y}$ denotes the forward scattering amplitude corresponding to the Fourier transform of the "dipole" correlator of lightlike Wilson lines in the fundamental representation \cite{Gelis:2010nm}, $\pi R_p^2$ is the effective transverse area of the proton~\cite{Ma:2014mri} and $\varphi_{p}$ is an unintegrated gluon distribution inside the proton:
\beq
\varphi_{p}(x_1,\vka)= \pi R_p^2 \frac{N_c k_{1\perp}^2}{4\alpha_s}\tilde{\mathcal{N}}_{Y}(x_1,\vka)\,,
\eeq 
where $\tilde{\mathcal{N}}_{Y}$ is the Fourier transform of a dipole correlator, but in this case with lightlike Wilson lines that live in the adjoint representation. 
These dipole forward scattering amplitudes, $\mathcal{N}_{Y}$ and $\tilde{\mathcal{N}}_{Y}$, are obtained by solving the running coupling Balitsky-Kovchegov (rcBK) equation \cite{Balitsky:1995ub,Kovchegov:1999yj} in momentum space, as a function of $x$, with McLerran--Venugopalan (MV) initial conditions \cite{McLerran:1993ni,McLerran:1993ka} specified at an initial large scale $x_0=0.01$~\cite{Albacete:2012xq}. For $x>0.01$, we employ an extrapolation of the solutions of the rcBK equation~\cite{Ma:2014mri} which is constrained by requiring that the corresponding integrated gluon distribution matches that in the collinear factorization framework.

We refer the reader to \cite{Kang:2013hta,Ma:2014mri}, and the references therein, for details of the derivation of these expressions. The novel feature here is the unwrapping (so to speak) of the helicity integrated expressions derived in \cite{Kang:2013hta} to extract the helicity dependent functions $\Gamma^{\kappa}_{ij}$ and $ {\cal G}^\kappa_{ij}$. The procedure is outlined in Appendix \ref{gamma_G_appendix}, where we provide the detailed expressions for these functions as well.

\section{Numerical results}
\label{num_res_sect}

We will now explicitly compute the expressions in Eqs.~(\ref{eq:dsktCO}) and (\ref{eq:dsktCS}) and use these to determine the angular distribution coefficients specifying $J/\psi$ polarization. In our parameter set for the numerical computations, we will set the charm mass to be $m_c=1.5\gev$--nearly one half of the $J/\psi$ mass. The value of the color singlet LDME is estimated using the value of the wavefunction at the origin in a potential model \cite{Eichten:1995ch}: $\mops=1.16/(2N_c) \gev^3$. For the color octet LDMEs, we employ the values obtained in Ref.~\cite{Chao:2012iv} by fitting NLO collinear factorized pQCD + NRQCD results to the Tevatron high $p_T$ prompt $J/\psi$ yields data: $\mopa=0.089\pm0.0098 \gev^3$, $\mopb=0.0030\pm0.0012 \gev^3$ and $\mopc/m_c^2=0.0056\pm0.0021\gev^3$. We will not use other sets of LDMEs extracted at NLO \cite{Gong:2012ug,Butenschoen:2011ks,Bodwin:2015iua} as they contain negative values for some of the LDMEs\footnote{Our impact factors, $\Gamma^{\kappa}_{ij}$, ${\cal G}^{\kappa}_{ij}$ are calculated at LO, so combining them with negative LDME may lead to negative cross sections.}.

The solution of the rcBK equation employs the code of Albacete et al. \cite{Albacete:2012xq} with MV initial conditions and the initial input parameters $\gamma=1$, $Q_{s0,\rm{proton}}^2=0.2\gev^2$, $\alpha_{fr}=0.5$ and $C=1$; these were determined from fits to the HERA DIS data \cite{Albacete:2012xq}. We have checked that our results for the angular coefficients, being ratios of cross-sections, are insensitive to the values of these parameters.
The theoretical errors we quote therefore are for the angular coefficients (collectively denoted henceforth as $\lambda$) and are obtained by varying the LDME values by their statistical uncertainties and by taking the minimal/maximal value of the obtained set.

\subsection{Spin density matrix elements in specific color channels}
\label{den_matrx_sect}

\begin{figure}
\begin{center}
\includegraphics[width=.47\textwidth]{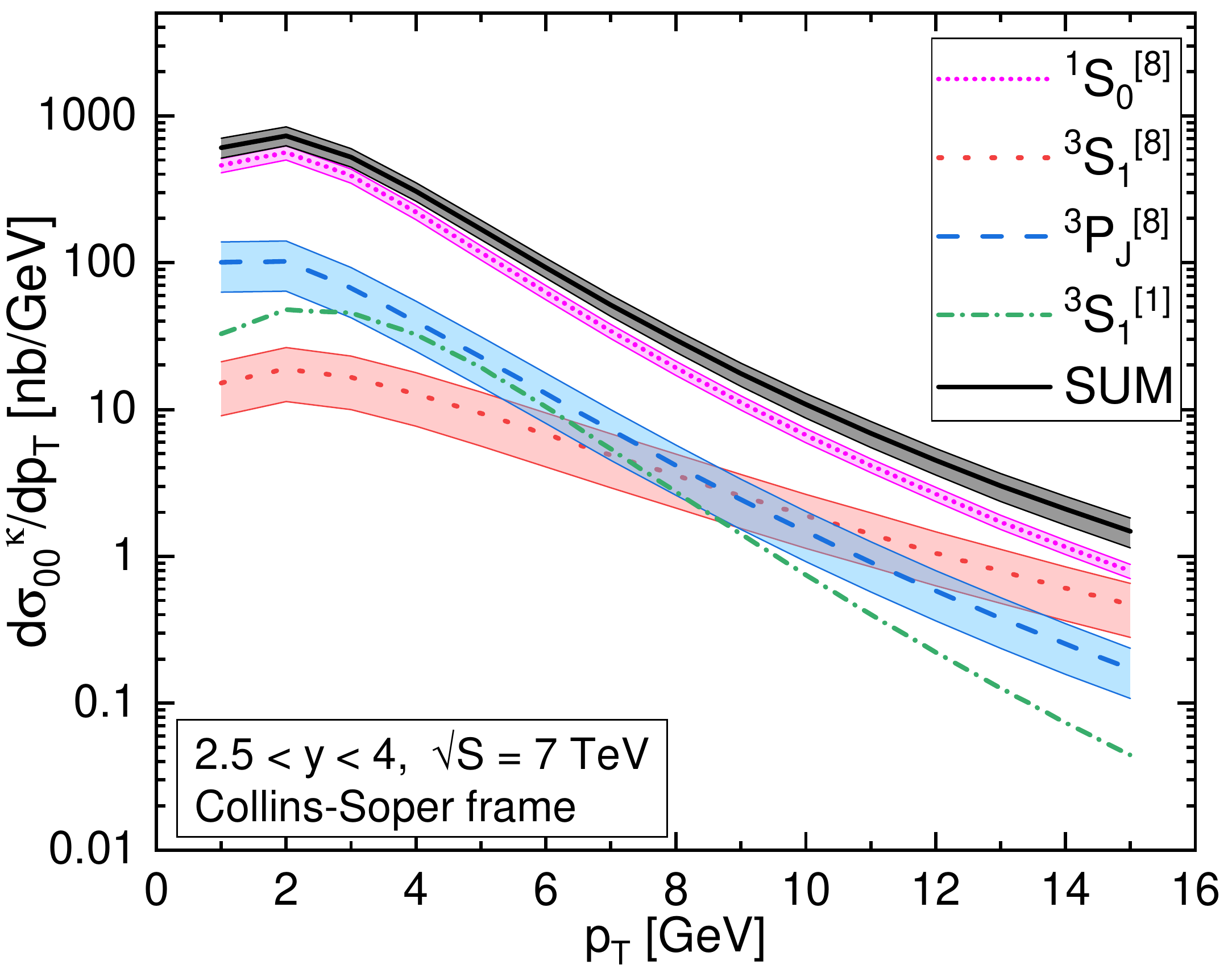}
\vspace{0.2cm}
\includegraphics[width=.47\textwidth]{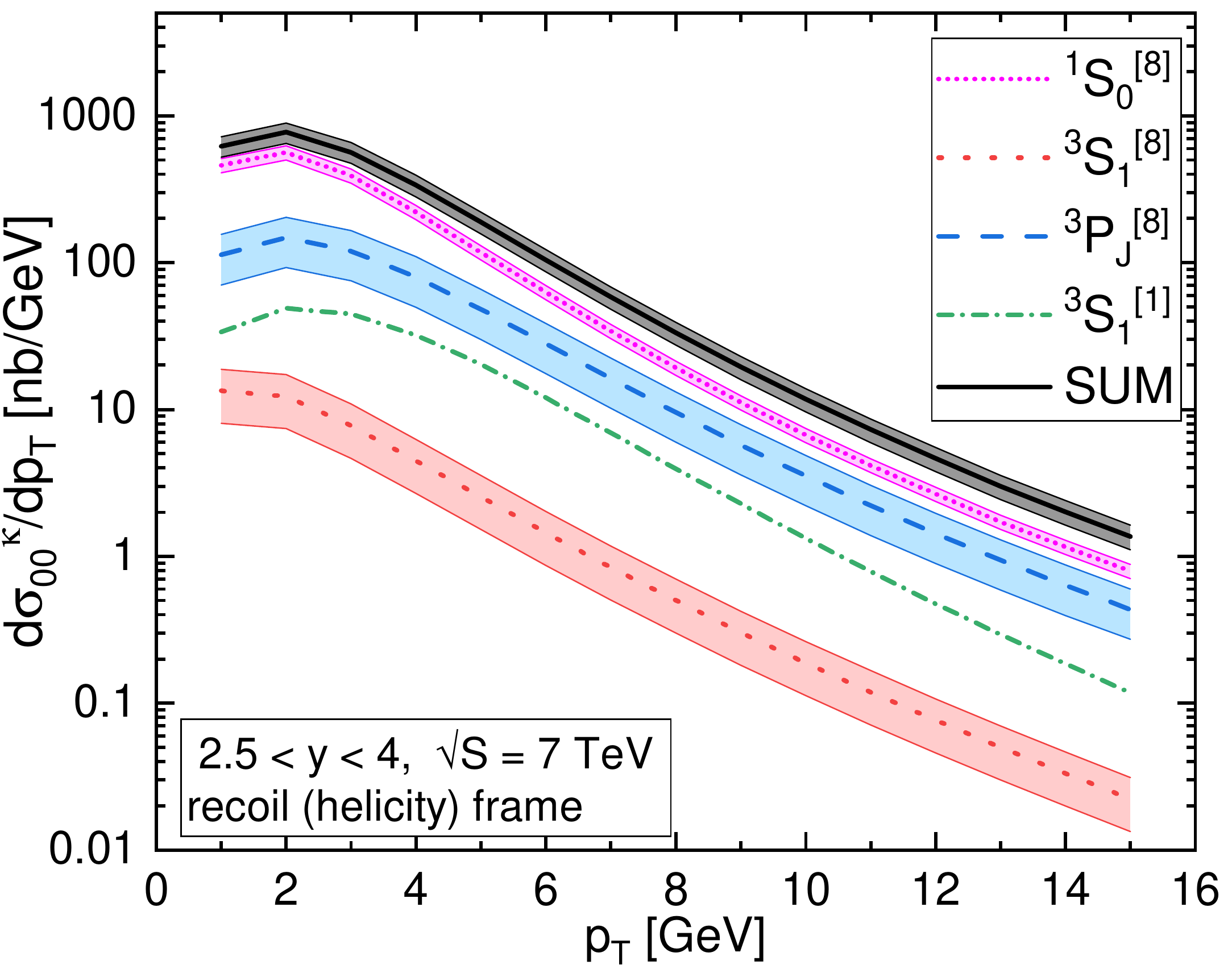}
\includegraphics[width=.47\textwidth]{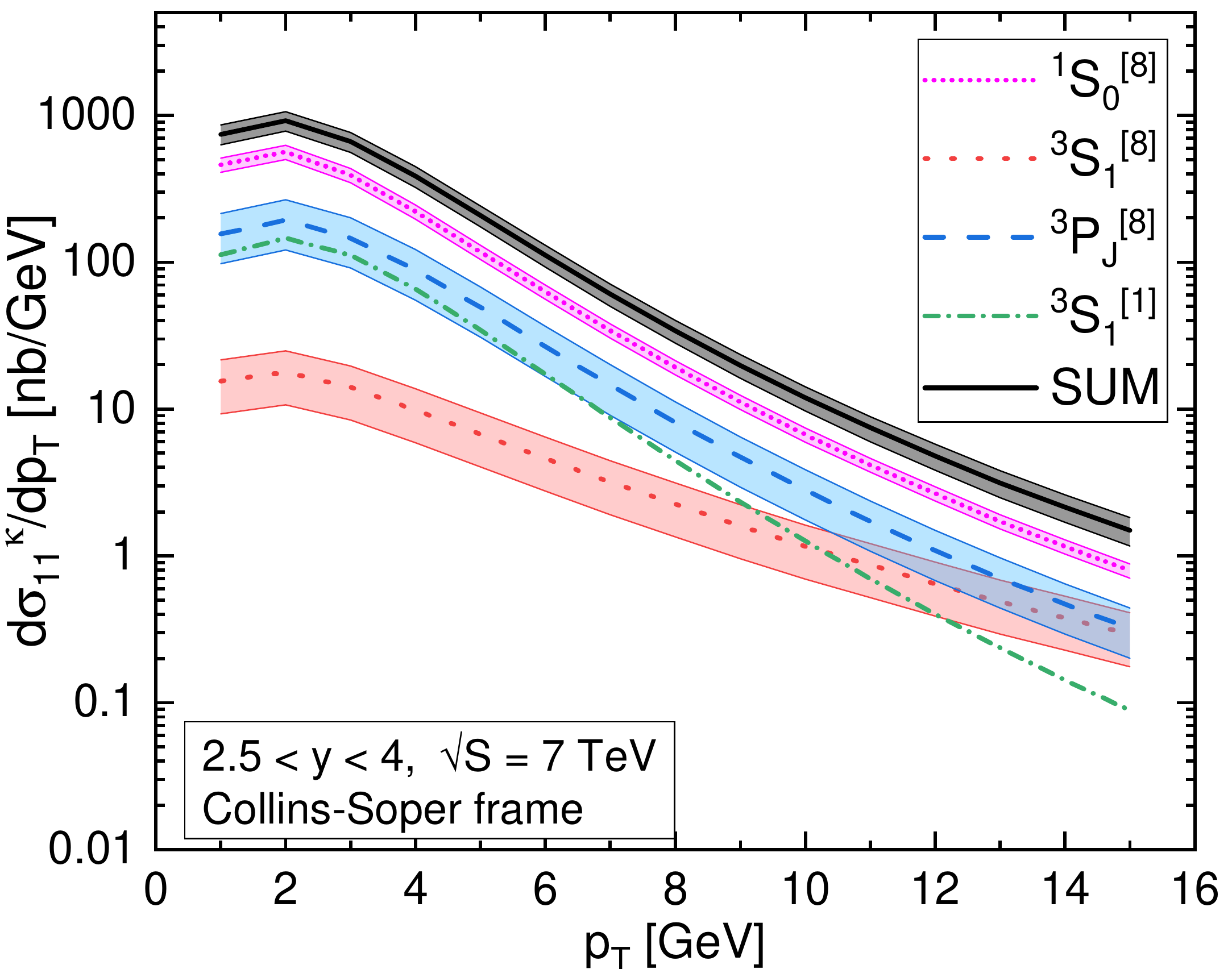}
\includegraphics[width=.47\textwidth]{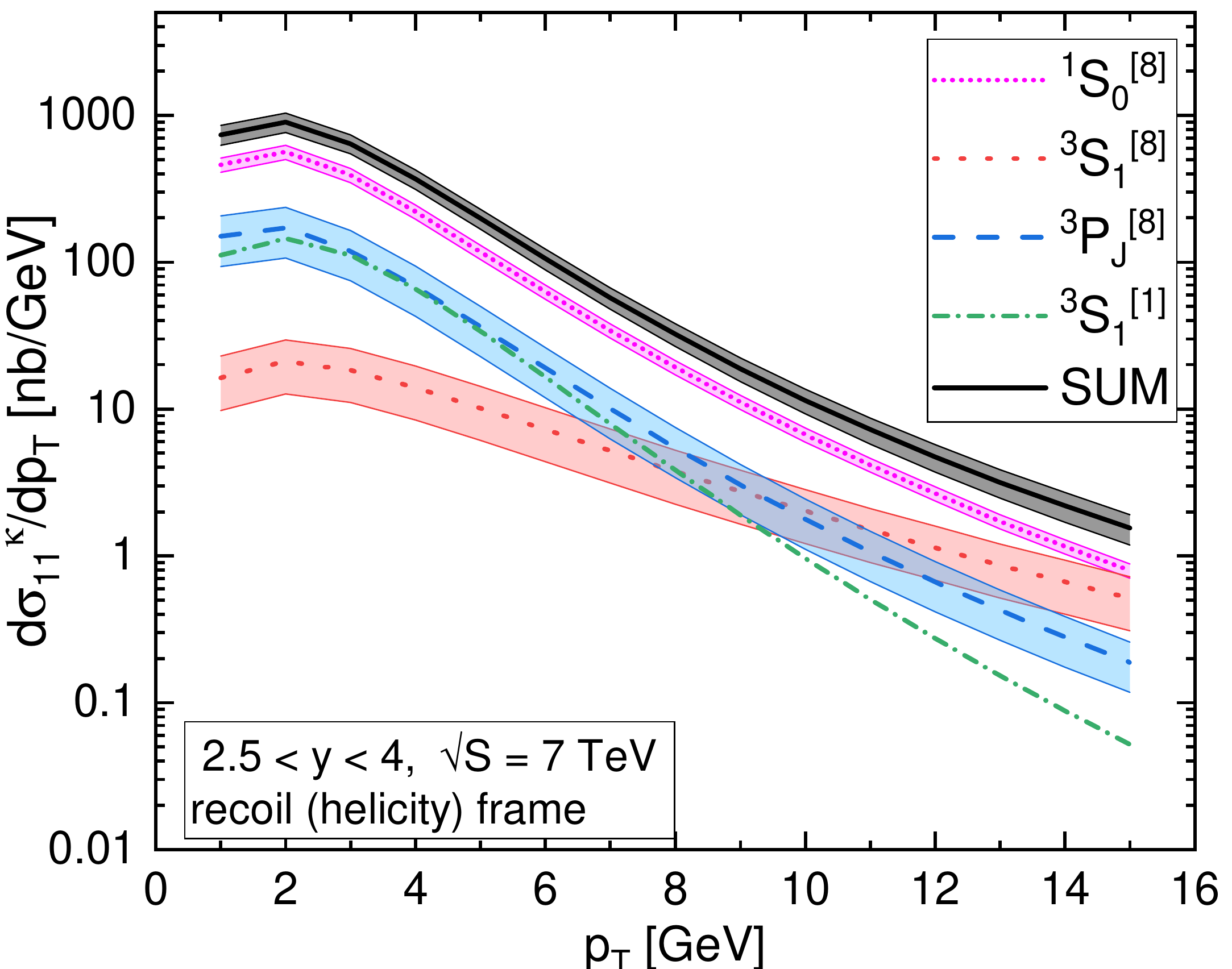}
\end{center}
\caption{The spin density matrix elements $d{\sigma}_{00}^\kappa$ (upper row) and $d{\sigma}_{11}^\kappa$ (lower row) in the Collins-Soper frame (left column) and in the recoil (right column) frame as functions of the $J/\psi$ transverse momentum $p_T$. The different curves represent contributions from different intermediate quantum states. "SUM" represents the sum over all states.
}
\label{sigma_mv_CS}
\end{figure}

We begin with a comparison of the contributions from different channels to the spin density matrix elements. In Figure \ref{sigma_mv_CS}, we plot the the matrix elements with $ij=\{00, 11\}$ for different quantum states. The rapidity interval is chosen to be $2.5\le Y \le 4$ and the center-of mass energy is $\sqrt{S}=7\tev$. One sees immediately that the $^1S_0^{[8]}$ state is the dominant channel for both matrix elements, as was also seen in the NLO collinear pQCD calculations \cite{Butenschoen:2012px}. The color singlet state $^3S_1^{[1]}$ contribution is similar to that from the $^3P_J^{[8]}$ state in the Collins-Soper frame; they are both larger than the $^3S_1^{[8]}$ state at low $p_T$ and then decrease rapidly with $p_T$ such that $^3S_1^{[8]}$ starts to become important at higher $p_T$. This is also the case for 
$\sigma_{11}$ in the recoil frame but $\sigma_{00}$ for the $^3S_1^{[8]}$ state decreases as fast as the other channels. A similar behavior can be seen in Figure 2 of \cite{Butenschoen:2012px}. This explains why in the in the recoil frame at high $p_T$ we have strong transverse polarization in the $^3S_1^{[8]}$ channel.

Even though the $^1S_0^{[8]}$ state is numerically dominant by far in the unpolarized cross-section in Eq.~(\ref{unpol_xsect}), it gives vanishing values for the polarization coefficients $\lamTheta$, $\lamPhi$ and $\lamThPh$ because the produced quark-antiquark pair has no spin and orbital momentum -- for further discussion, see \cite{Ma:2015yka}. We can write the helicity SDCs in this state as 
\bear
d\sigma^{\, ^1S_0^{[8]}}_{ij}=\begin{cases}
\frac{1}{3} d\sigma^{ ^1S_0^{[8]}} \ &\text{ if }ij=00,++, \text{ or } --\,,\\
0 \ &\text{ in other cases}\,,
\end{cases}
\enar
where $d\sigma^{ ^1S_0^{[8]}}$ is the unpolarized cross-section for this state. Knowing that $^1S_0^{[8]}$ dominates in the diagonal matrix elements $ij=\{11,00, -1-1\}$, one should expect a suppression of the polarization coefficients in Eq.~(\ref{lam_Defin}).

\subsection{Results for the $\lambda$ polarization coefficients in CGC+NRQCD}

\begin{figure}
\begin{center}
\includegraphics[width=.47\textwidth]{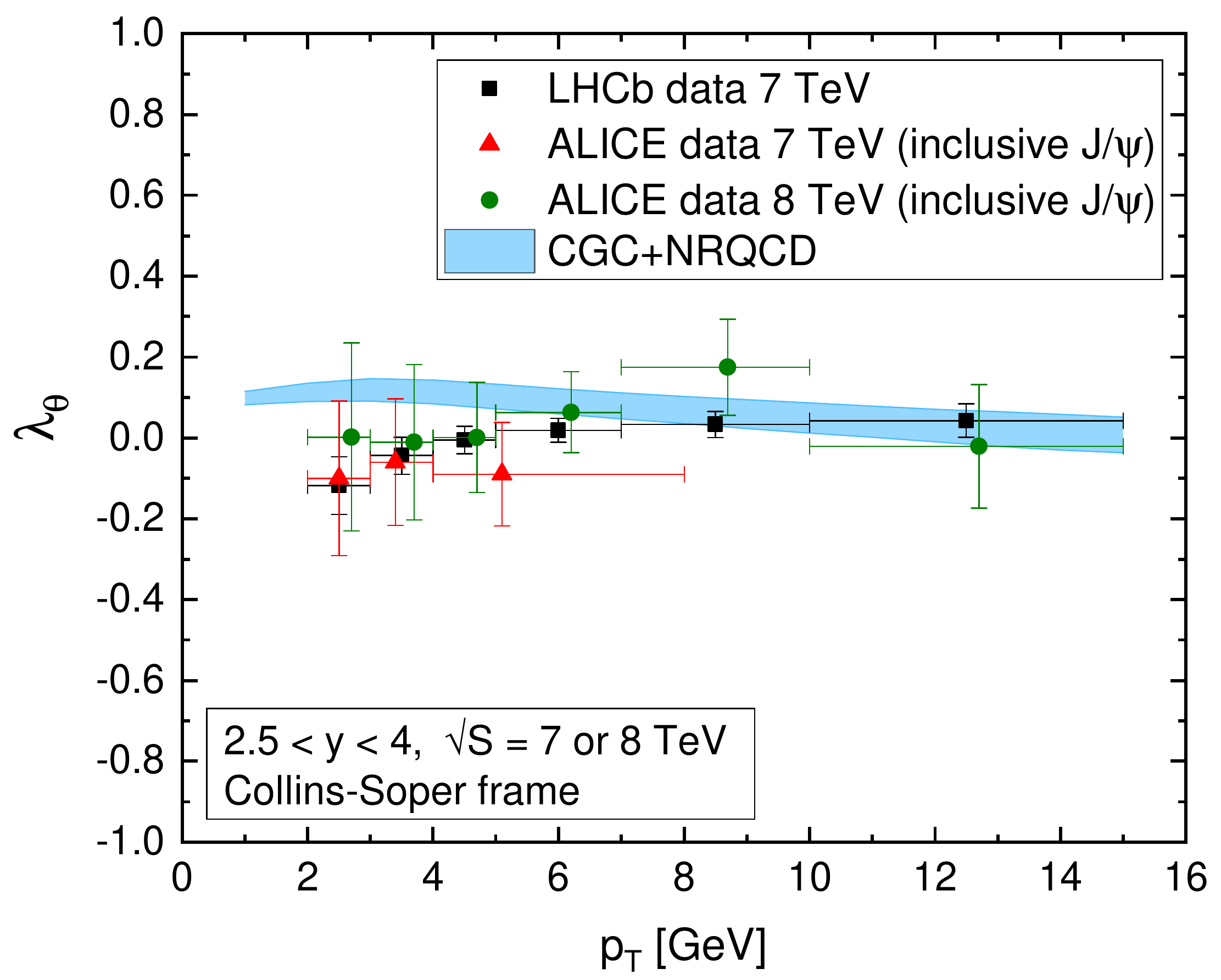}
\vspace{0.2cm}
\includegraphics[width=.47\textwidth]{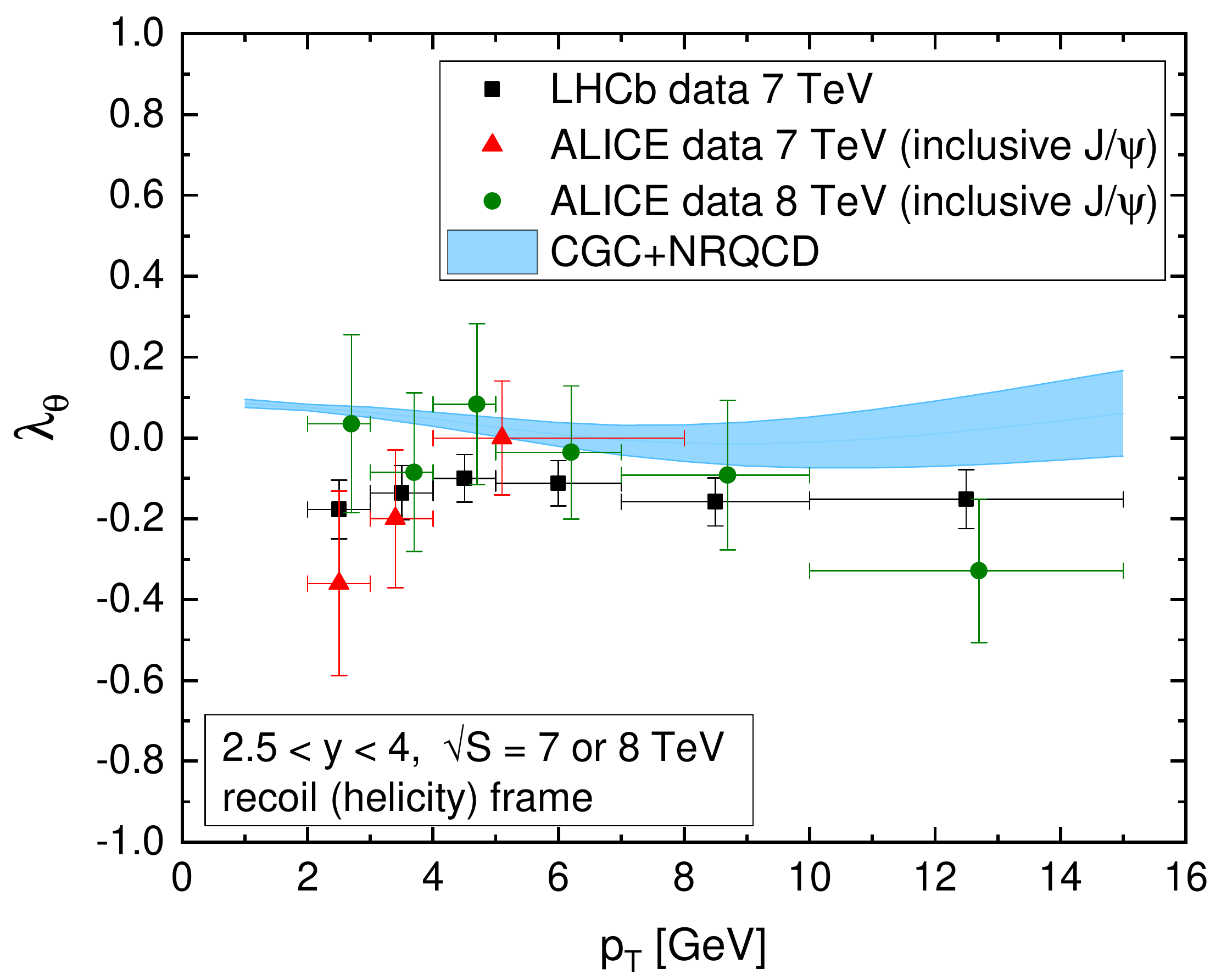}
\vspace{0.2cm}
\includegraphics[width=.47\textwidth]{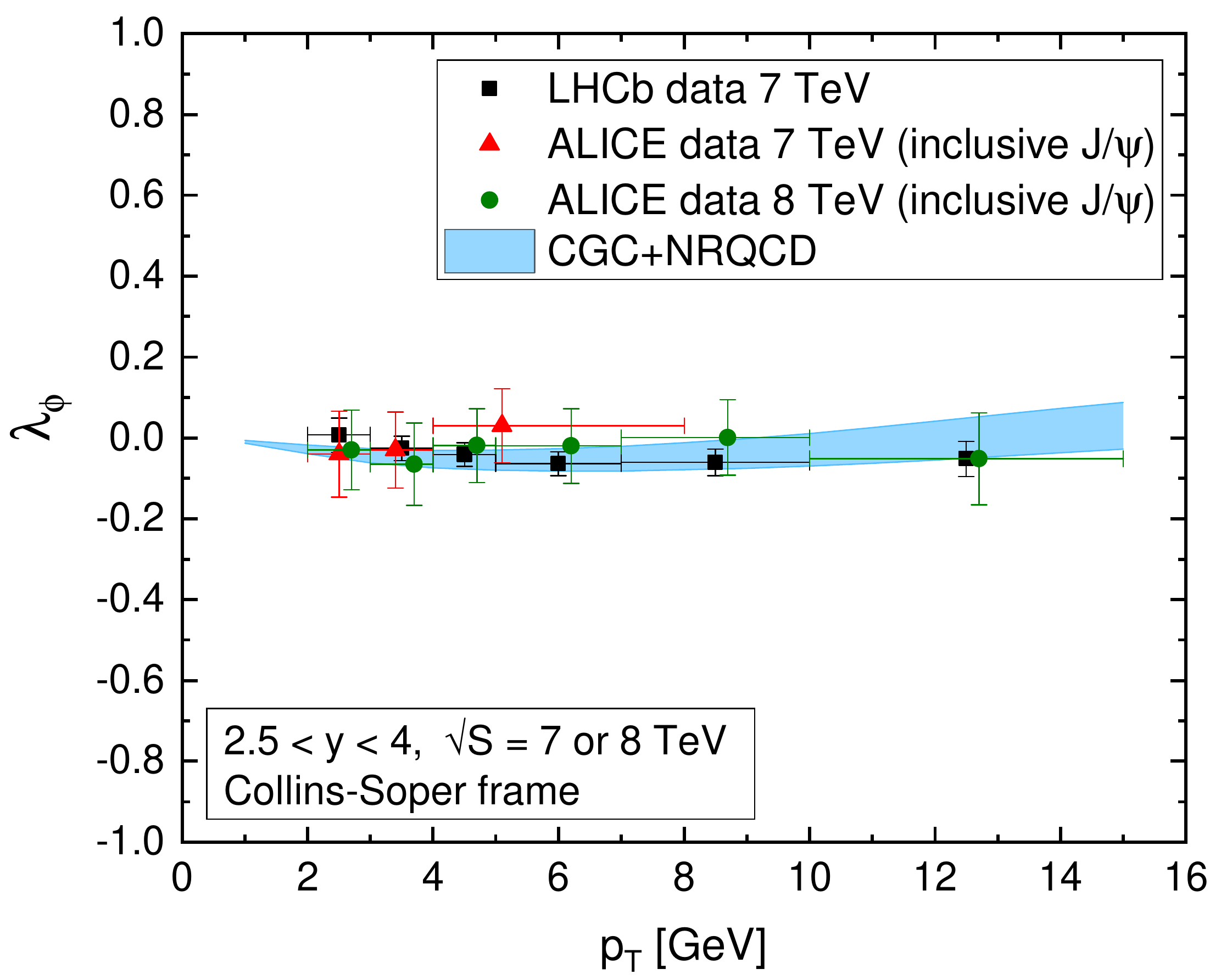}
\includegraphics[width=.47\textwidth]{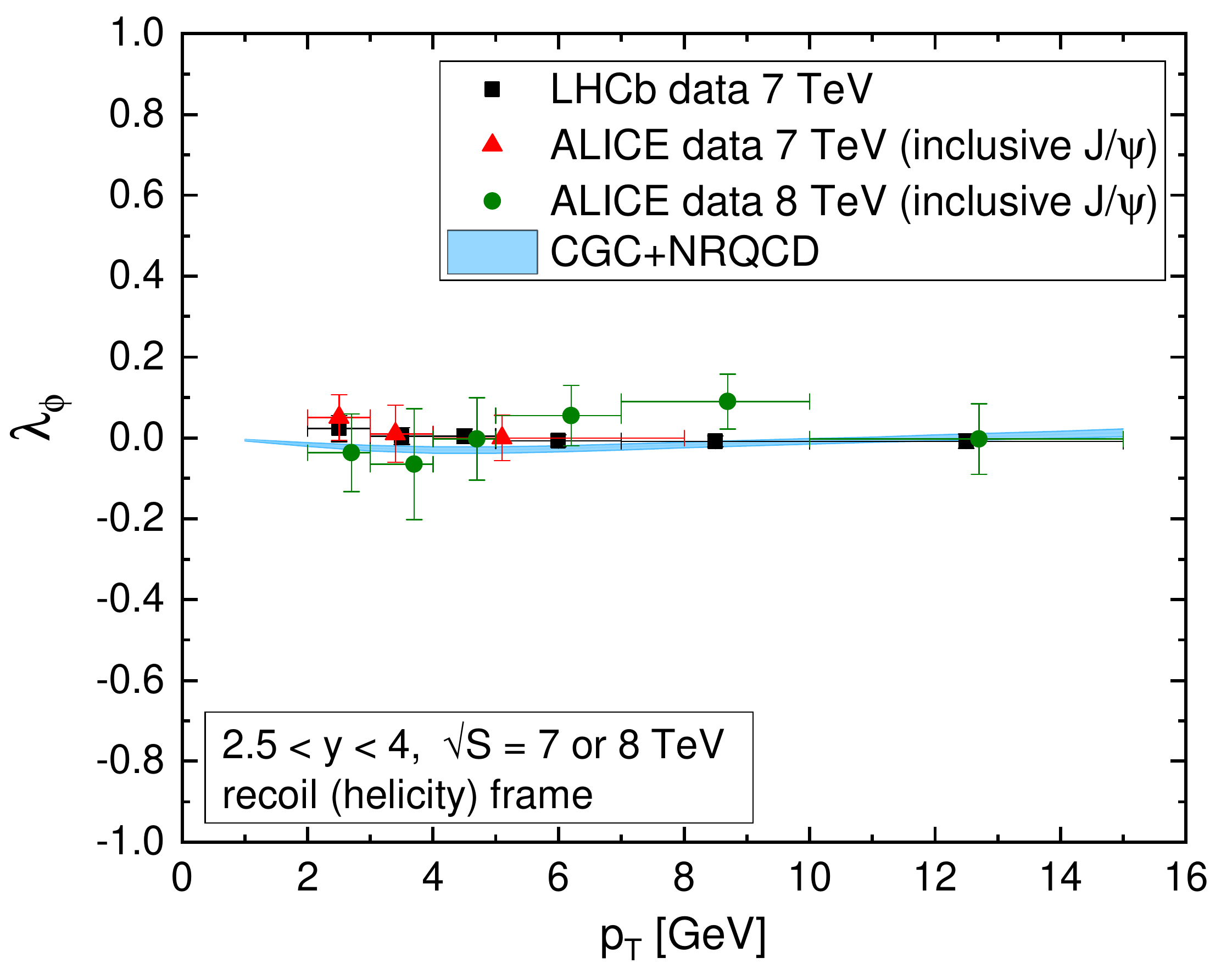}
\includegraphics[width=.47\textwidth]{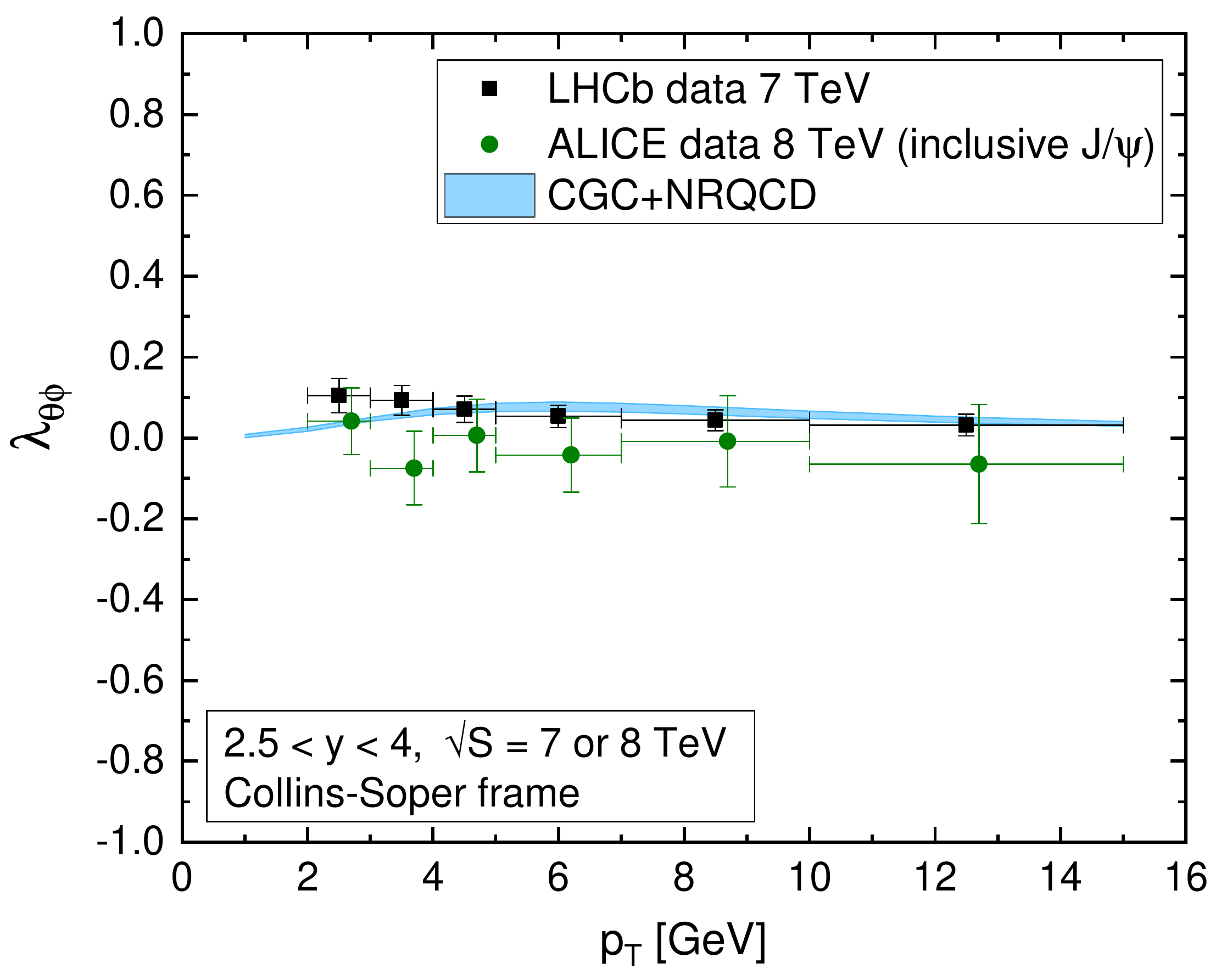}
\includegraphics[width=.47\textwidth]{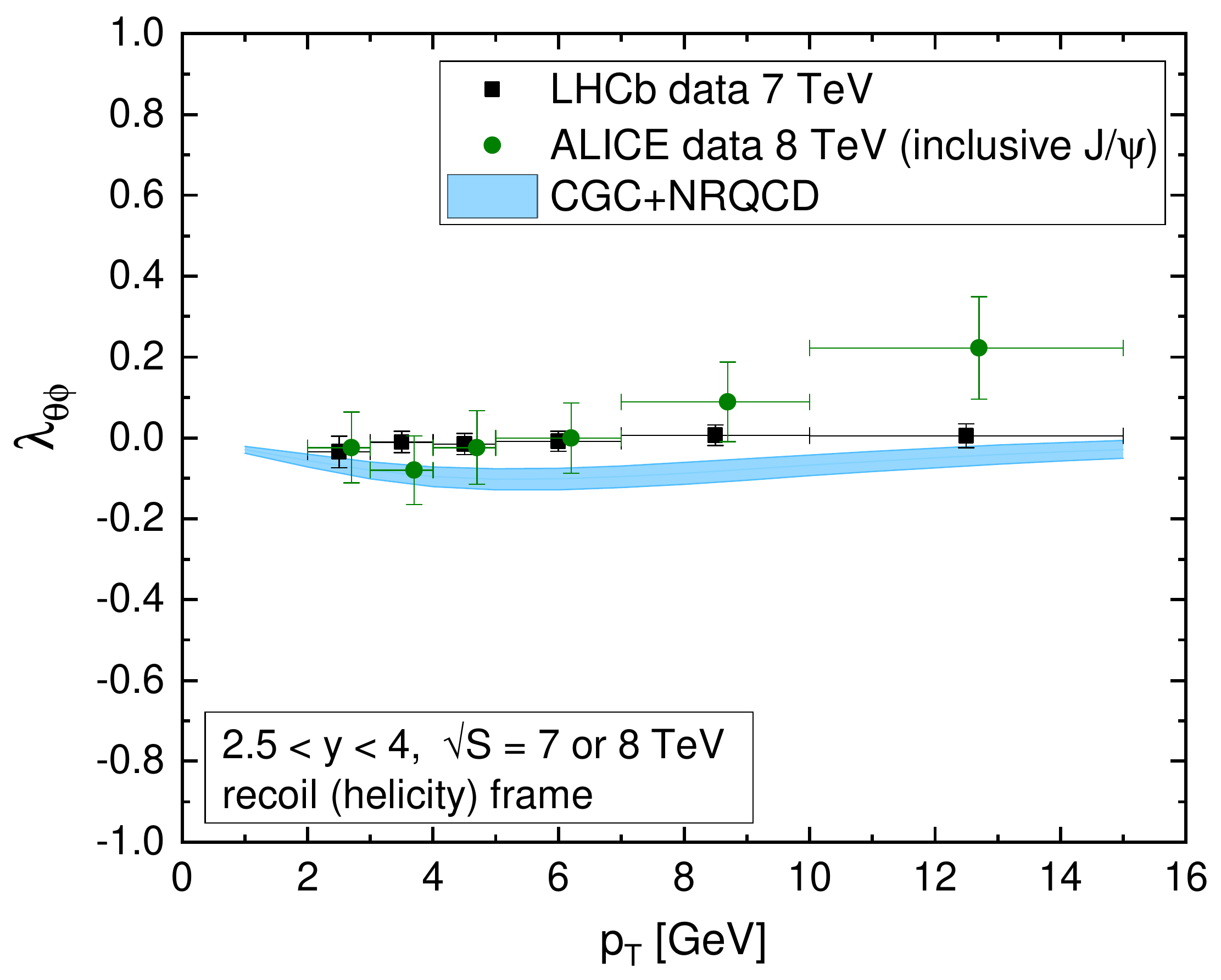}
\end{center}
\caption{The angular distribution coefficients $\lamTheta$ (first row), $\lamPhi$ (second row) and $\lamThPh$ (third row) in the Collins-Soper frame (left column) and in the recoil frame (right column) as functions of the $J/\psi$ transverse momentum $p_T$. Data are from the LHCb experiment at $7\tev$ \cite{Aaij:2013nlm}, the ALICE experiment at $7\tev$ \cite{Abelev:2011md} and ALICE at $8\tev$ \cite{Acharya:2018uww}, all in the $2.5<y<4$ rapidity window. Note that the ALICE data are for inclusive $J/\psi$ production, containing contributions from $B$-meson decays.
}
\label{lambda_mv_LHCb_ALICE_7_8}
\end{figure}

In Figure \ref{lambda_mv_LHCb_ALICE_7_8}, we show results for all three angular distribution coefficients $\lambda$ compared to data in the Collins--Soper frame (left column) and in the recoil frame (right column). The following data were used\footnote{We have checked that the results for $7\tev$ and $8\tev$ are very close to each other -they are indistinguishable in the plots.}: LHCb at $7\tev$ \cite{Aaij:2013nlm}, ALICE at $7\tev$ \cite{Abelev:2011md} and ALICE at $8\tev$ \cite{Acharya:2018uww}. Both LHCb and ALICE measured angular coefficients in the rapidity window $2.5<y<4$. Note that ALICE data are obtained for inclusive $J/\psi$ production, so they also contain contributions from $B$-meson decays. However the contribution from $B$-meson decays are on the order of a few percent at low $p_T$ \cite{Aaij:2011jh}, so we can neglect them here\footnote{The good agreement between the LHCb data for prompt production and the ALICE data for inclusive production affirms this statement.}. 

The polarization parameter $\lamTheta$ is measured to be close to zero, indicating that the $J/\psi$ are mostly unpolarized. At small $p_T$, our results prefer a small transverse polarization ($\lamTheta>0$). The data on the other hand seem to prefer a weak longitudinal polarization ($\lamTheta < 0$) albeit it should be noted that there is considerable variation between the experiments, with LHCb and ALICE $7\tev$ data showing negative central values and ALICE $8\tev$ data showing positive values. The data are consistent with each other to $1\sigma$ accuracy. At higher transverse momentum ($p_T \gtrsim 6\gev$), our results agree with data within two standard deviations. We note that the agreement of our theory results with the data in this kinematic region is significantly better than two of the three the collinearly factorized NLO pQCD+NRQCD computations; it is however close to the results of \cite{Chao:2012iv}. The compilations shown in \cite{Aaij:2013nlm} and \cite{Acharya:2018uww} comparing NLO pQCD+NRQCD and color singlet model results to data demonstrate that there is considerable variation between the data for $\lamTheta$ and those NLO pQCD+NRQCD theory results. 

For the $\lamPhi$ coefficient, we obtain very good agreement with data; the data are within the CGC+NRQCD theory band for both frames. For $\lamThPh$, we obtain very good agreement with the LHCb data in the Collins--Soper frame. Our results are higher than the ALICE $8\tev$ data though within $1\sigma$ accuracy; we note that there is tension between the LHCb and ALICE data at this level of accuracy. In the recoil frame, the ALICE data are well described except for high $p_T$, where data are systematically above the CGC+NRQCD predictions. In contrast, the agreement with the LHCb data is good at the higher $p_T$ while we are slightly below at low $p_T$. 

The take away message here is that the experimental values, as well as our theory results, for $\lamPhi$ and $\lamThPh$, are consistent with zero. Our description of data for $\lamPhi$ and $\lamThPh$ is significantly better than the NLO pQCD+NRQCD calculations of \cite{Butenschoen:2012px}: in fact, both the color singlet model and NLO pQCD+NRQCD approaches predict polarization coefficients with absolute values significantly larger than measured at LHCb and ALICE. A similar conclusion can be drawn for results obtained by a second group performing NLO pQCD+NRQCD analyses \cite{Gong:2012ug}. While, as noted, the third group \cite{Chao:2012iv,Shao:2014yta} obtain a good description of data for $\lamTheta$ in the recoil frame for $p_T>5\gev$, no results are provided for the $\lamPhi$ and $\lamThPh$ coefficients in this frame, or for any of the angular coefficients in the Collins--Soper frame.

\begin{figure}
\begin{center}
\includegraphics[width=.47\textwidth]{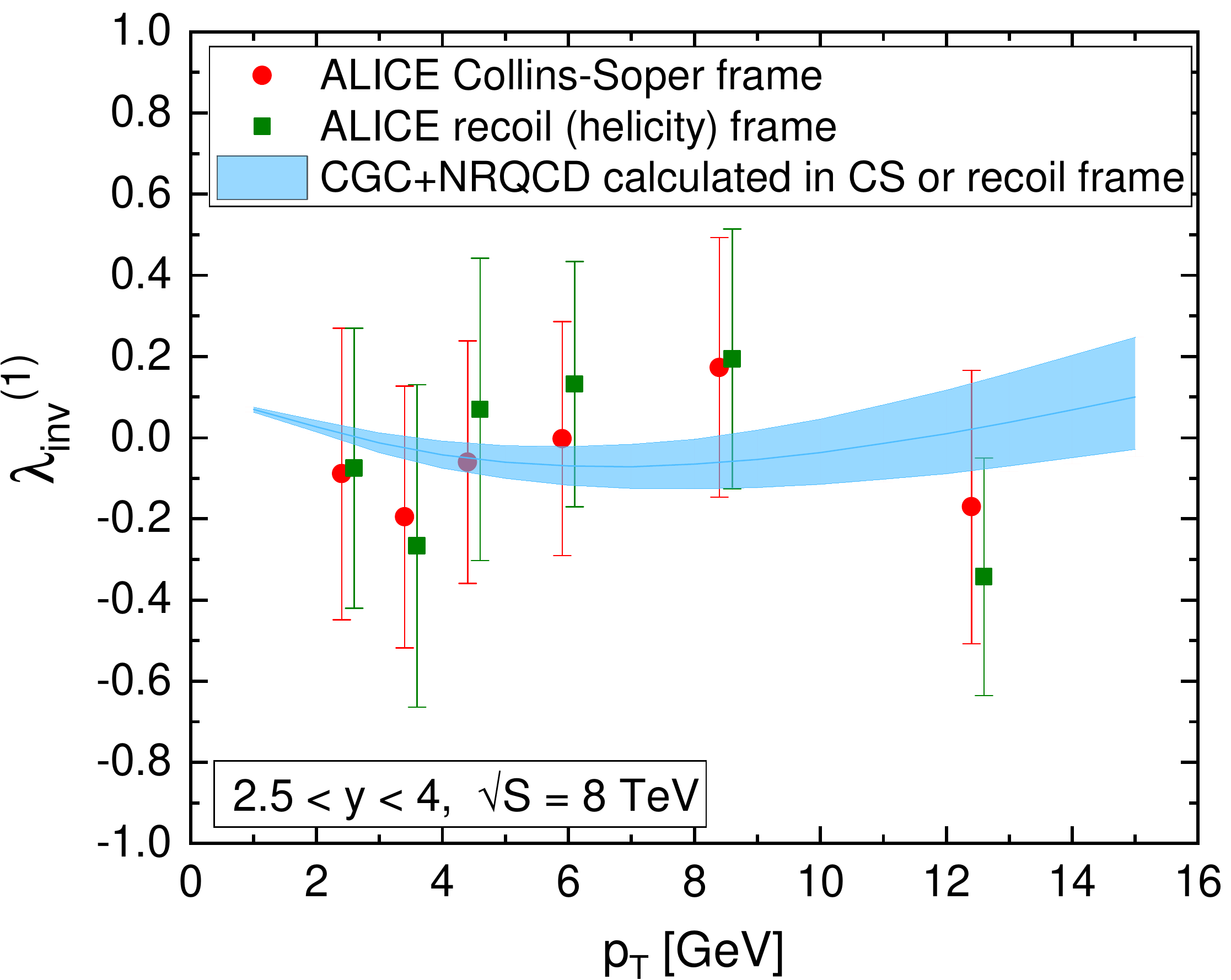}
\includegraphics[width=.47\textwidth]{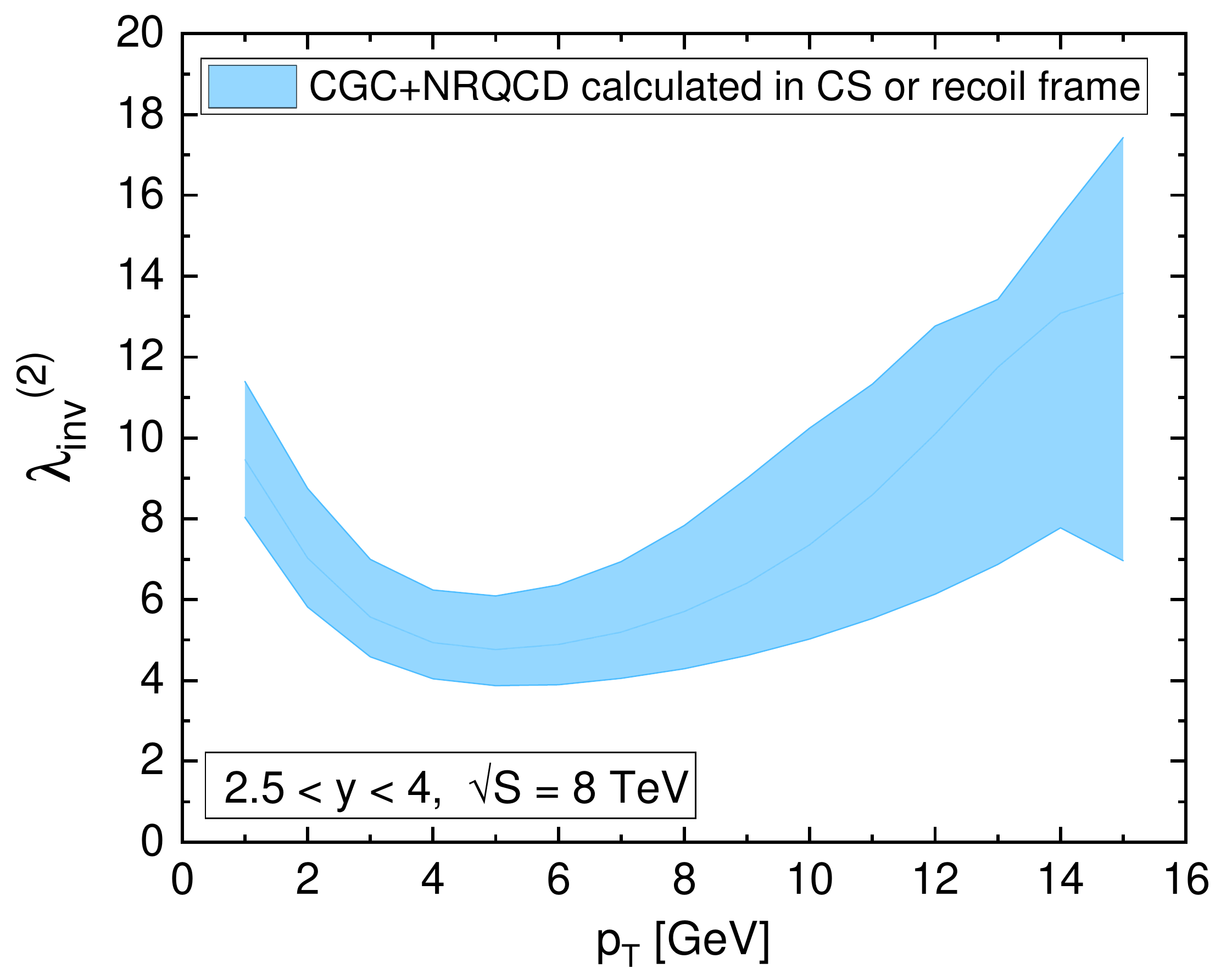}
\end{center}
\caption{The frame independent angular coefficients $\lamINV^{(1)}$ (left plot) and $\lamINV^{(2)}$ (right plot) as functions of the $J/\psi$'s transverse momentum $p_T$. The data are from the ALICE experiment at $8\tev$ \cite{Acharya:2018uww}. The experimental values of $\lamINV^{(1)}$ calculated using the Collins-Soper and recoil frames (represented by red and green points respectively) coincide with each other. 
}
\label{lambda_inv_ALICE}
\end{figure}

In Figure \ref{lambda_inv_ALICE}, we show our results for the frame invariant quantities $\lamINV$ defined in Eq.~(\ref{lam_inv_definition}), which are compared to the ALICE data \cite{Acharya:2018uww} at $8\tev$. Once again, one observes that CGC+NRQCD provides a good description of $\lamINV^{(1)}$ though indeed the error bars in the data are considerable. We also provide predictions for the $\lamINV^{(2)}$ coefficient that was proposed in \cite{Palestini:2010xu}. Since this coefficient is sensitive to all three of the frame-dependent polarization coefficients $\lamTheta$, $\lamPhi$, $\lamThPh$, we believe it may provide an additional constraint both for theoretical studies and experimental measurements.

\begin{figure}
\begin{center}
\includegraphics[width=.47\textwidth]{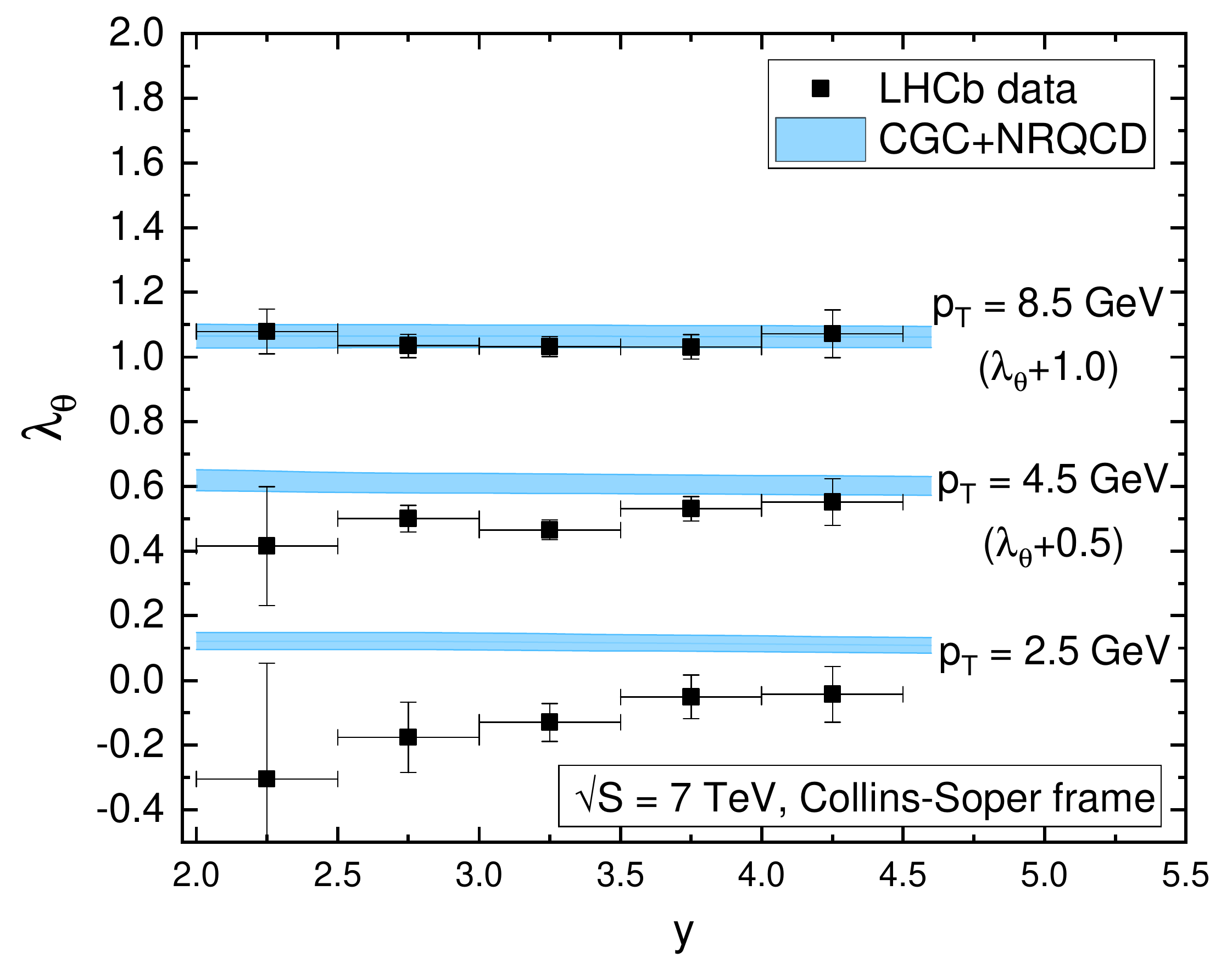}
\vspace{0.2cm}
\includegraphics[width=.47\textwidth]{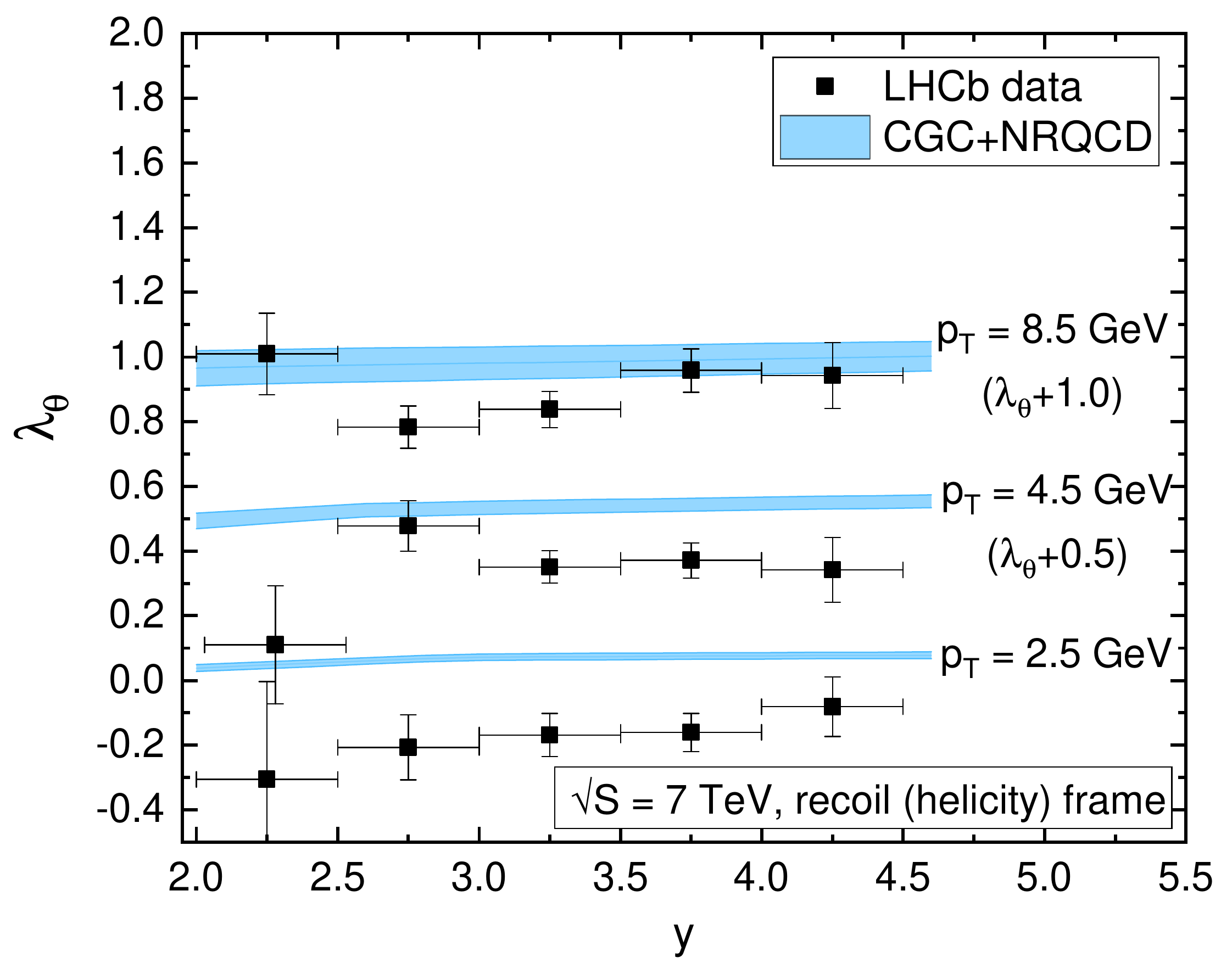}
\vspace{0.2cm}
\includegraphics[width=.47\textwidth]{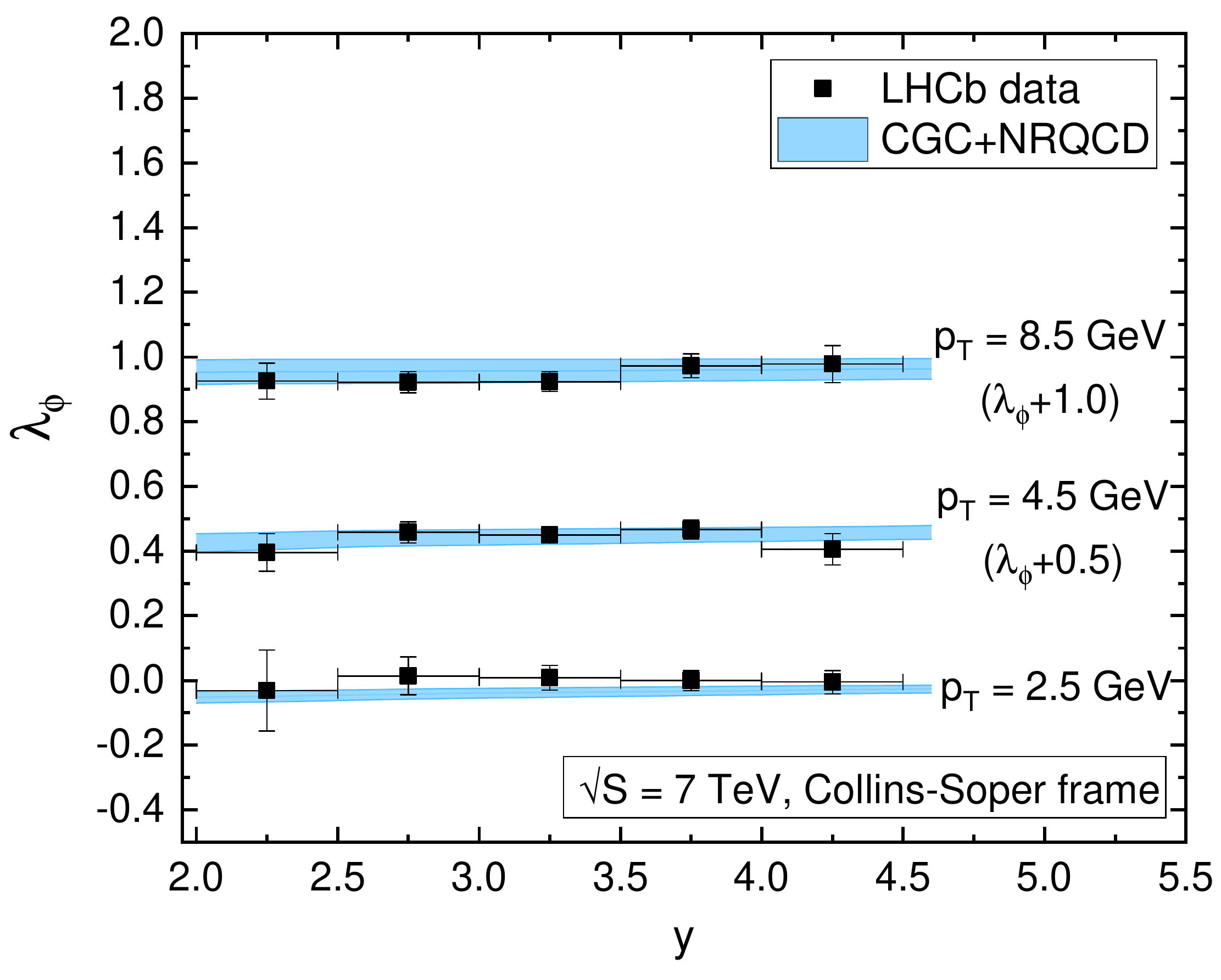}
\includegraphics[width=.47\textwidth]{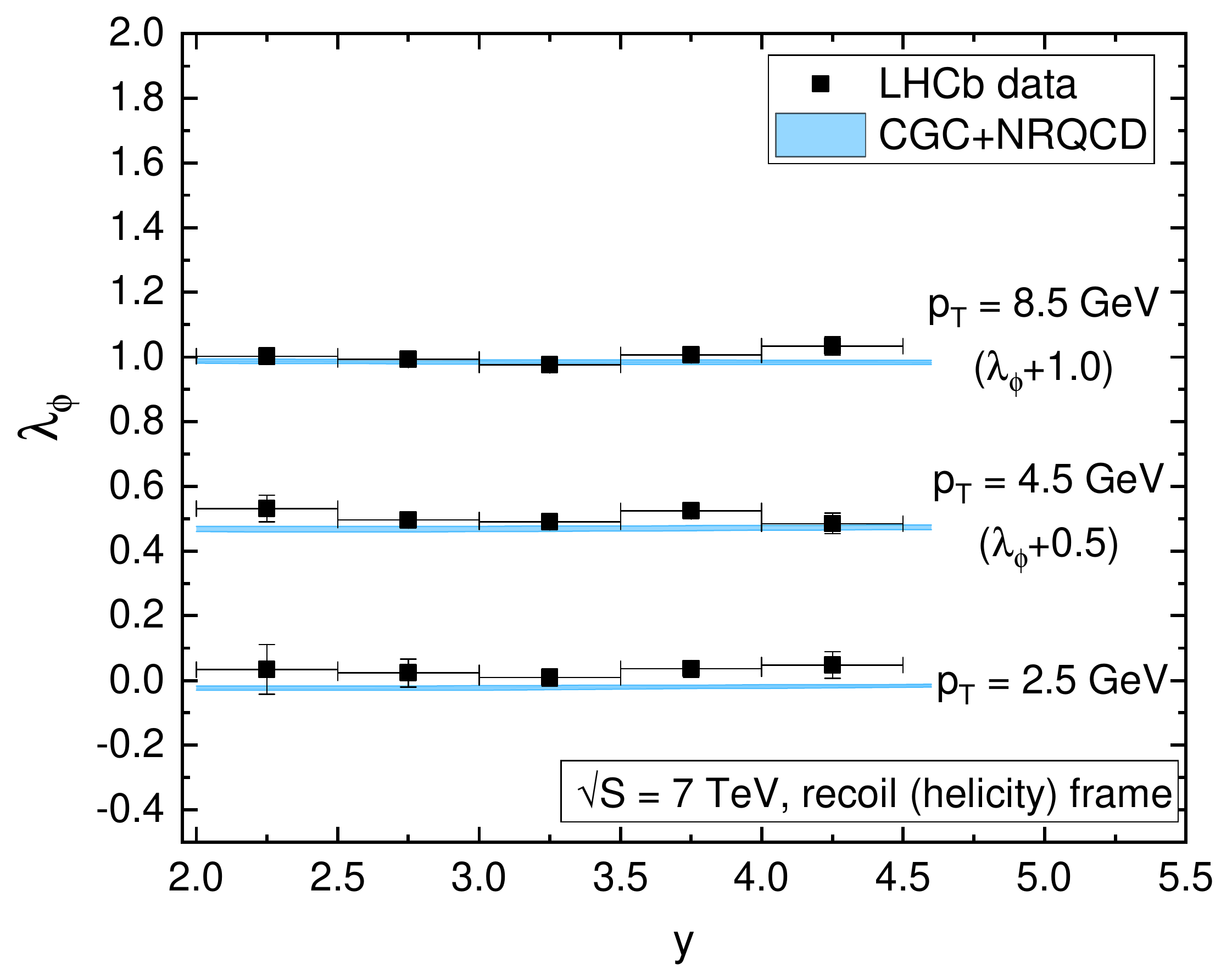}
\includegraphics[width=.47\textwidth]{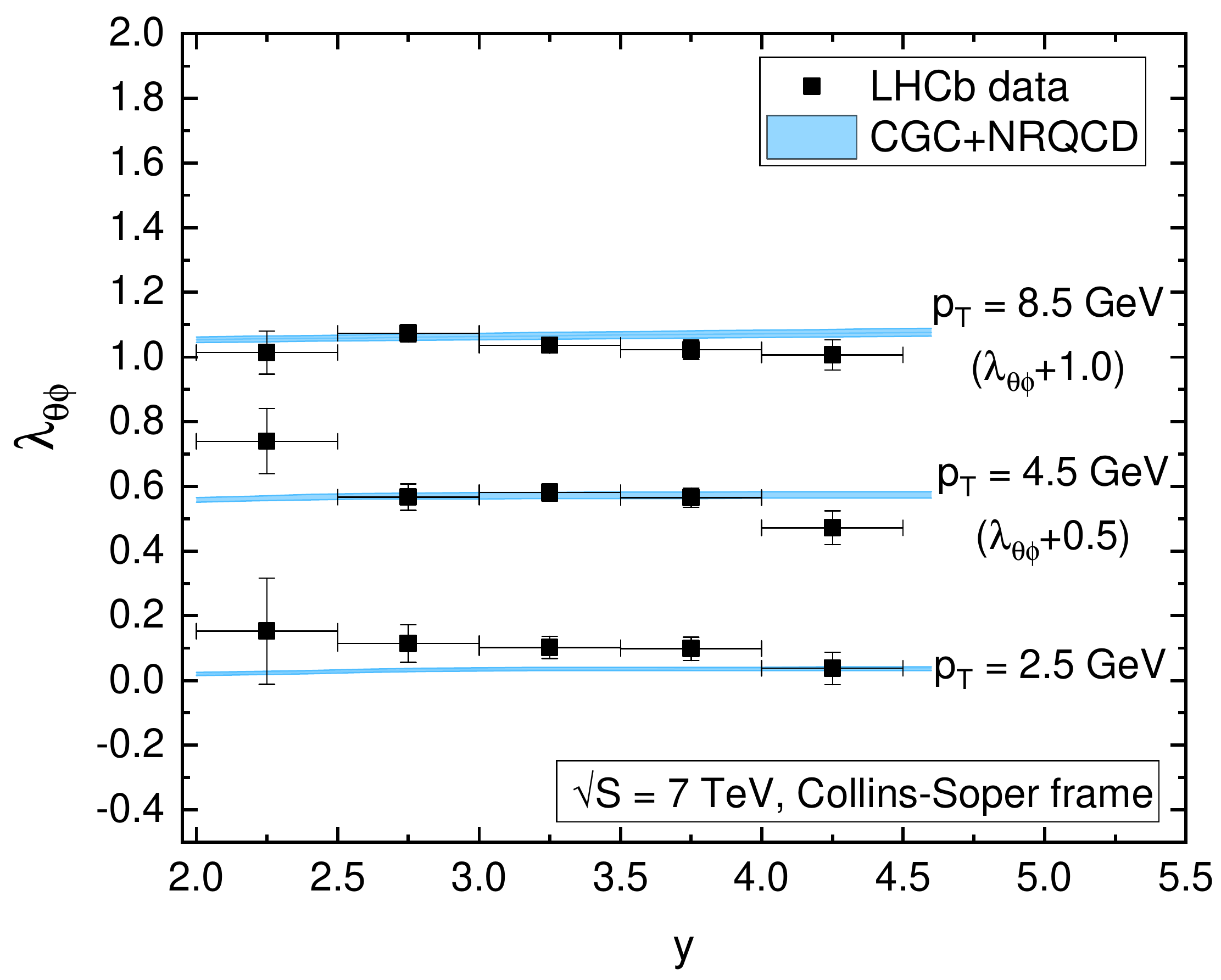}
\includegraphics[width=.47\textwidth]{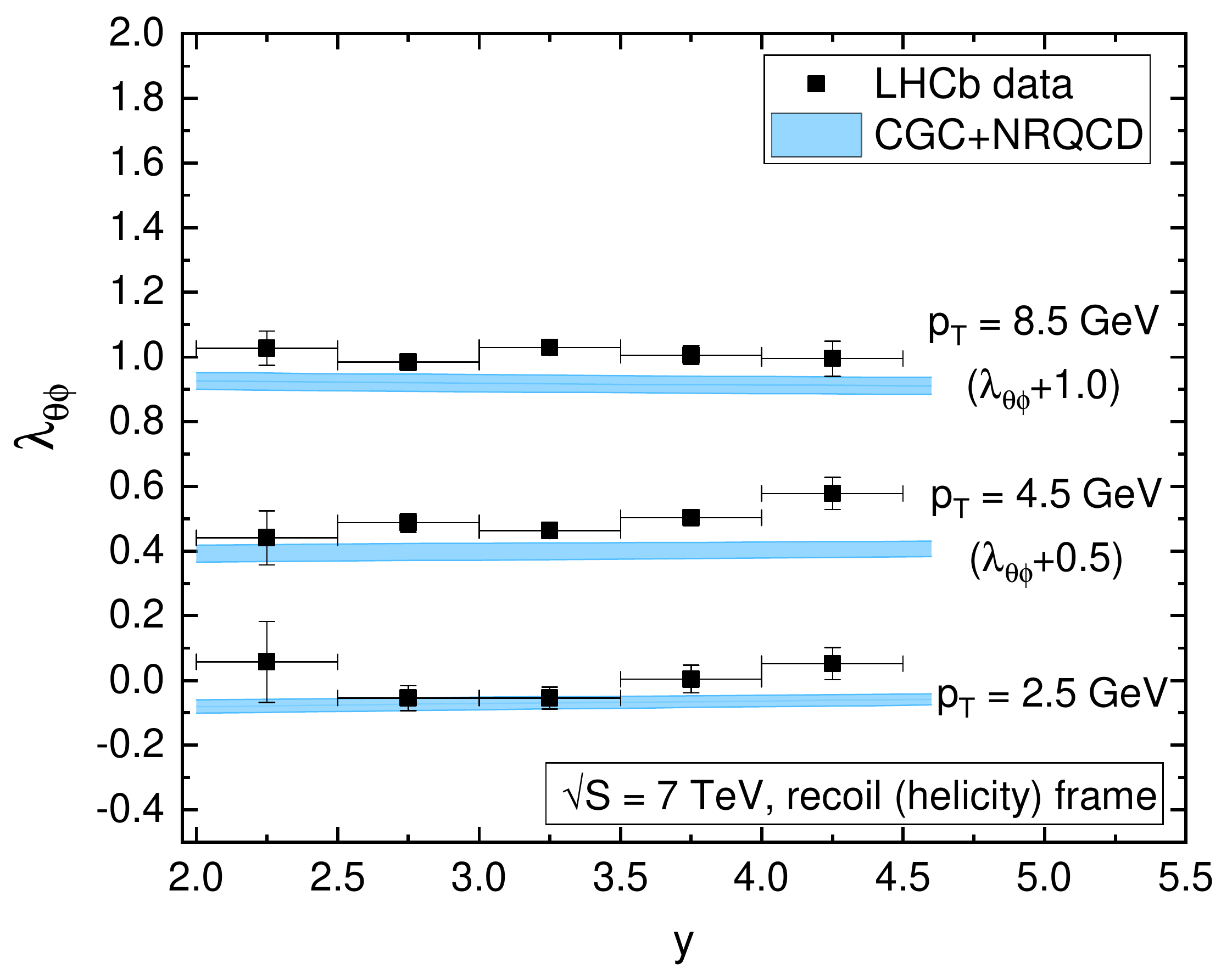}
\end{center}
\caption{The angular coefficients $\lamTheta$ (first row), $\lamPhi$ (second row) and $\lamThPh$ (third row) in the Collins-Soper frame (left column) and recoil frame (right column) as functions of the $J/\psi$'s rapidity $y$, plotted for three transverse momentum values $p_T=2.5, \ 4.5, \ 8.5 \gev$. Some values of $\lambda$ were shifted for better visibility: by 0.5 for $p_T=4.5\gev$ and by 1.0 for $p_T=8.5\gev$. Data are from the LHCb experiment at $7\tev$ \cite{Aaij:2013nlm}.
}
\label{lambda_mv_LHCb_y_dependence}
\end{figure}

\begin{figure}
\begin{center}
\includegraphics[width=.47\textwidth]{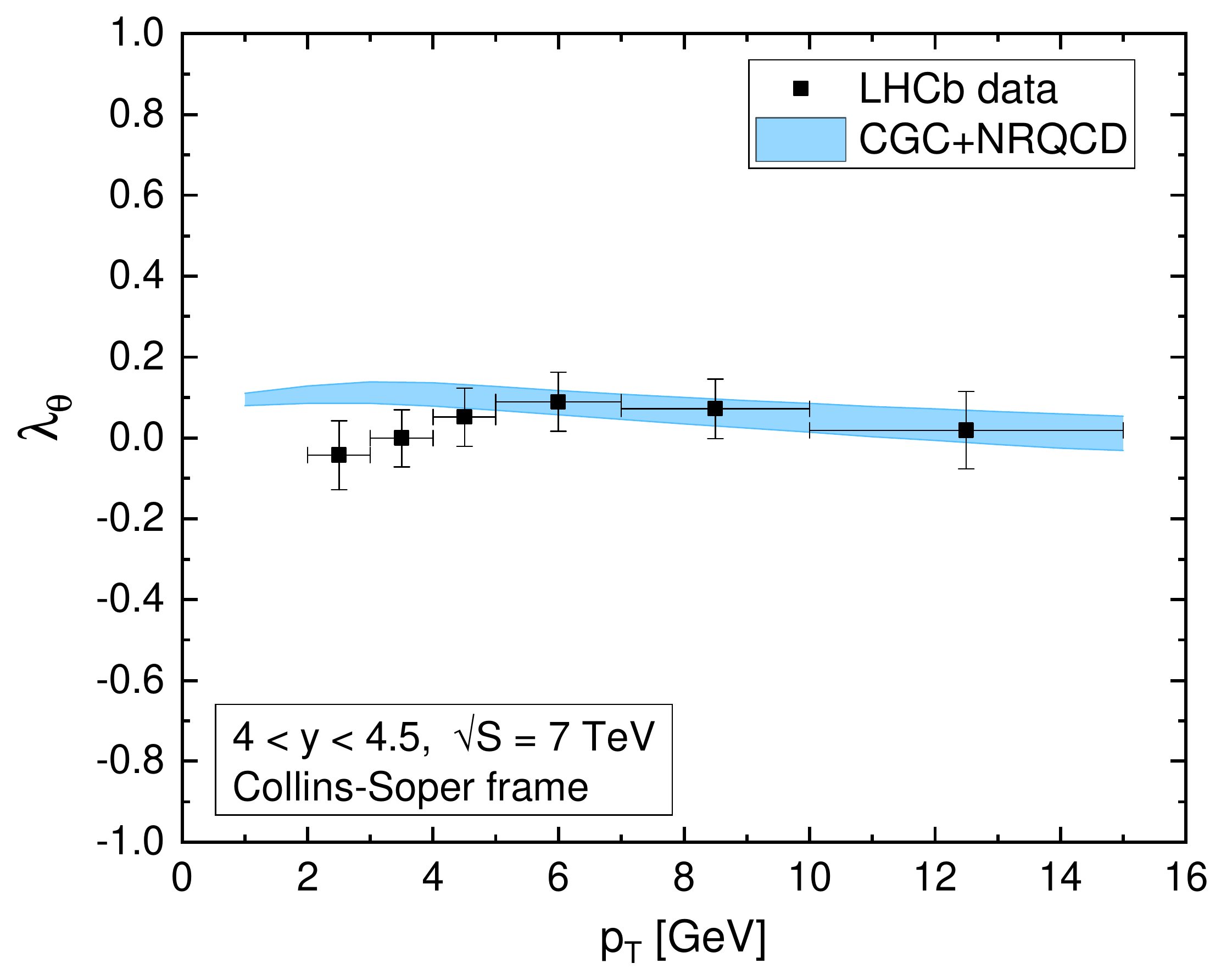}
\vspace{0.2cm}
\includegraphics[width=.47\textwidth]{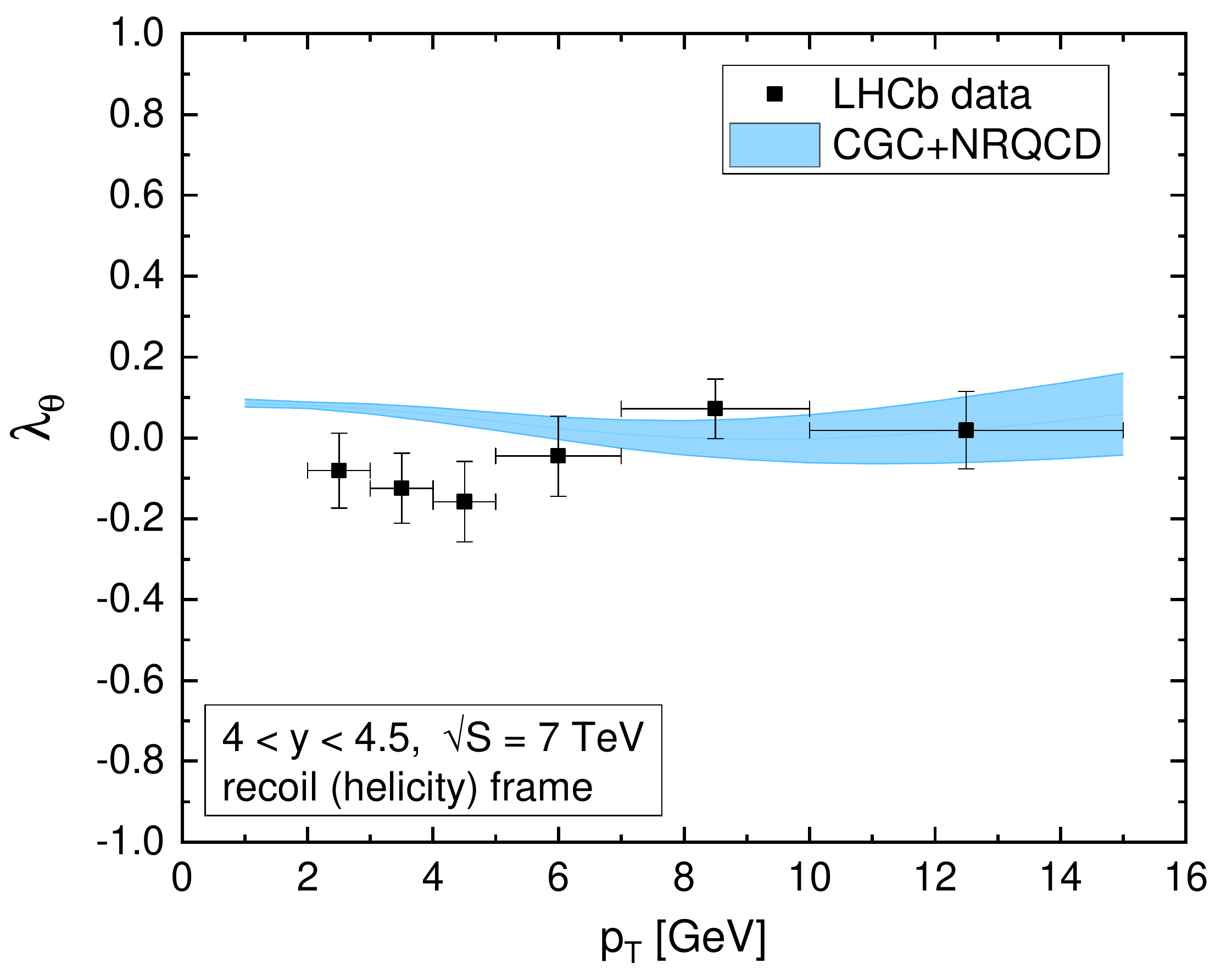}
\vspace{0.2cm}
\includegraphics[width=.47\textwidth]{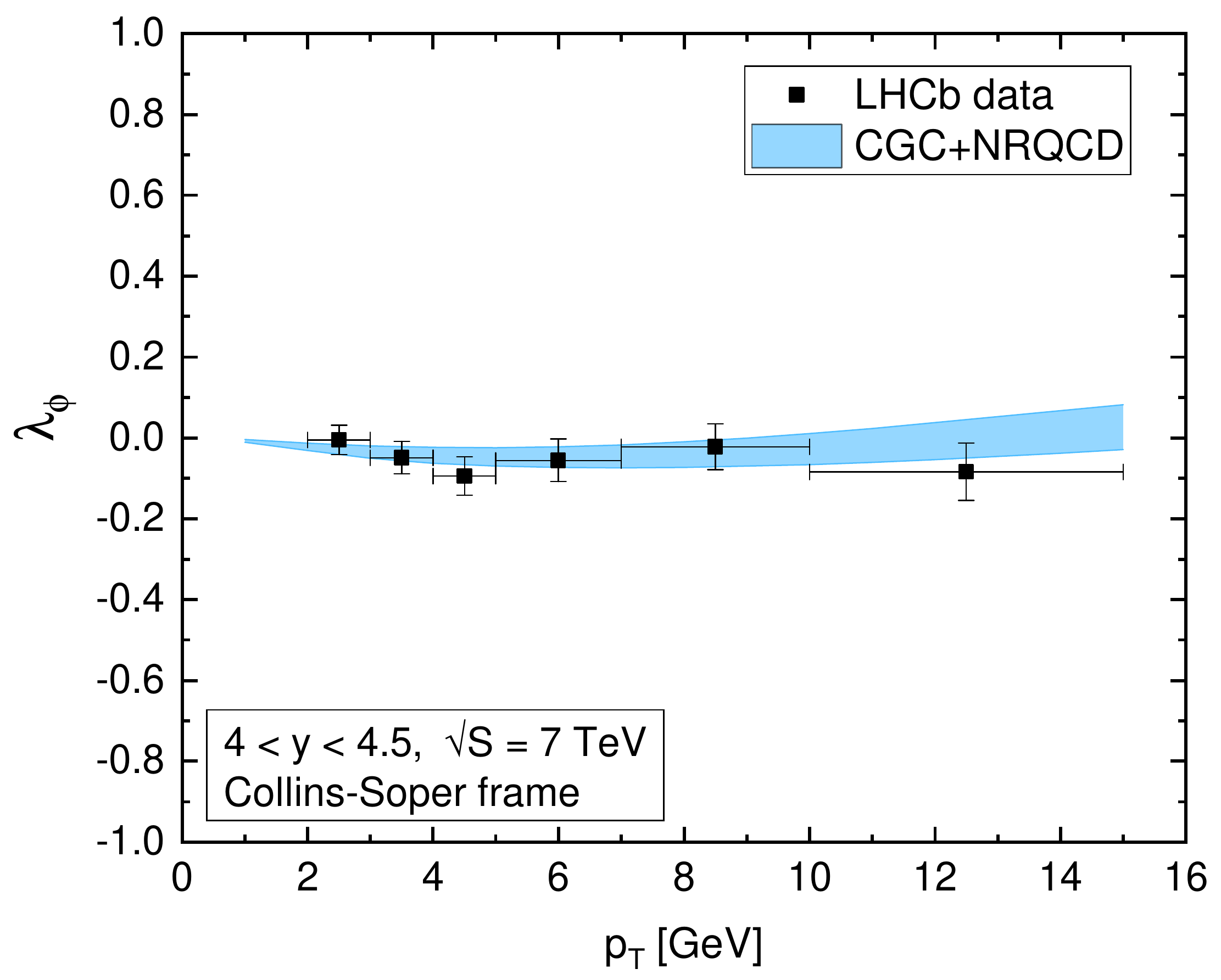}
\includegraphics[width=.47\textwidth]{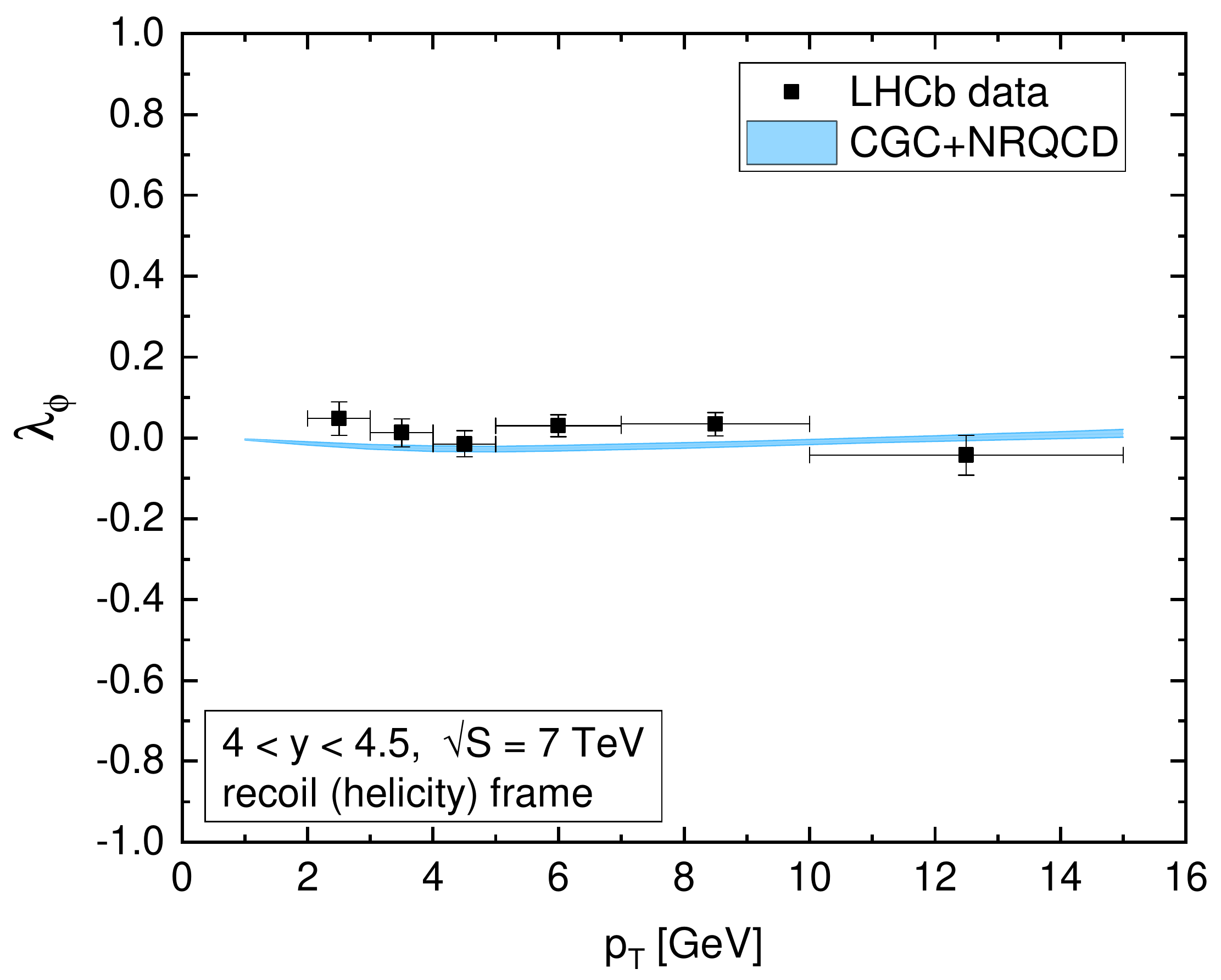}
\includegraphics[width=.47\textwidth]{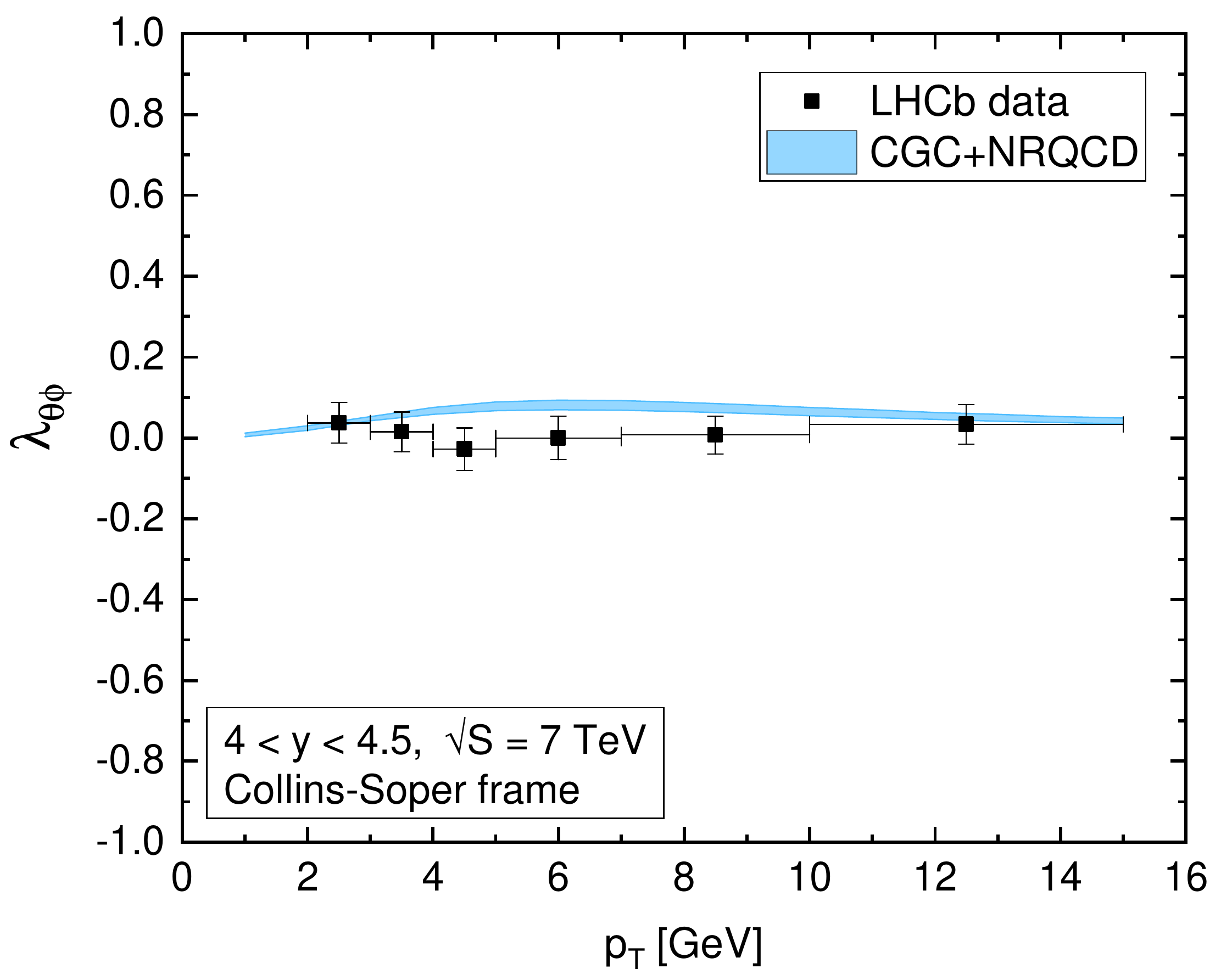}
\includegraphics[width=.47\textwidth]{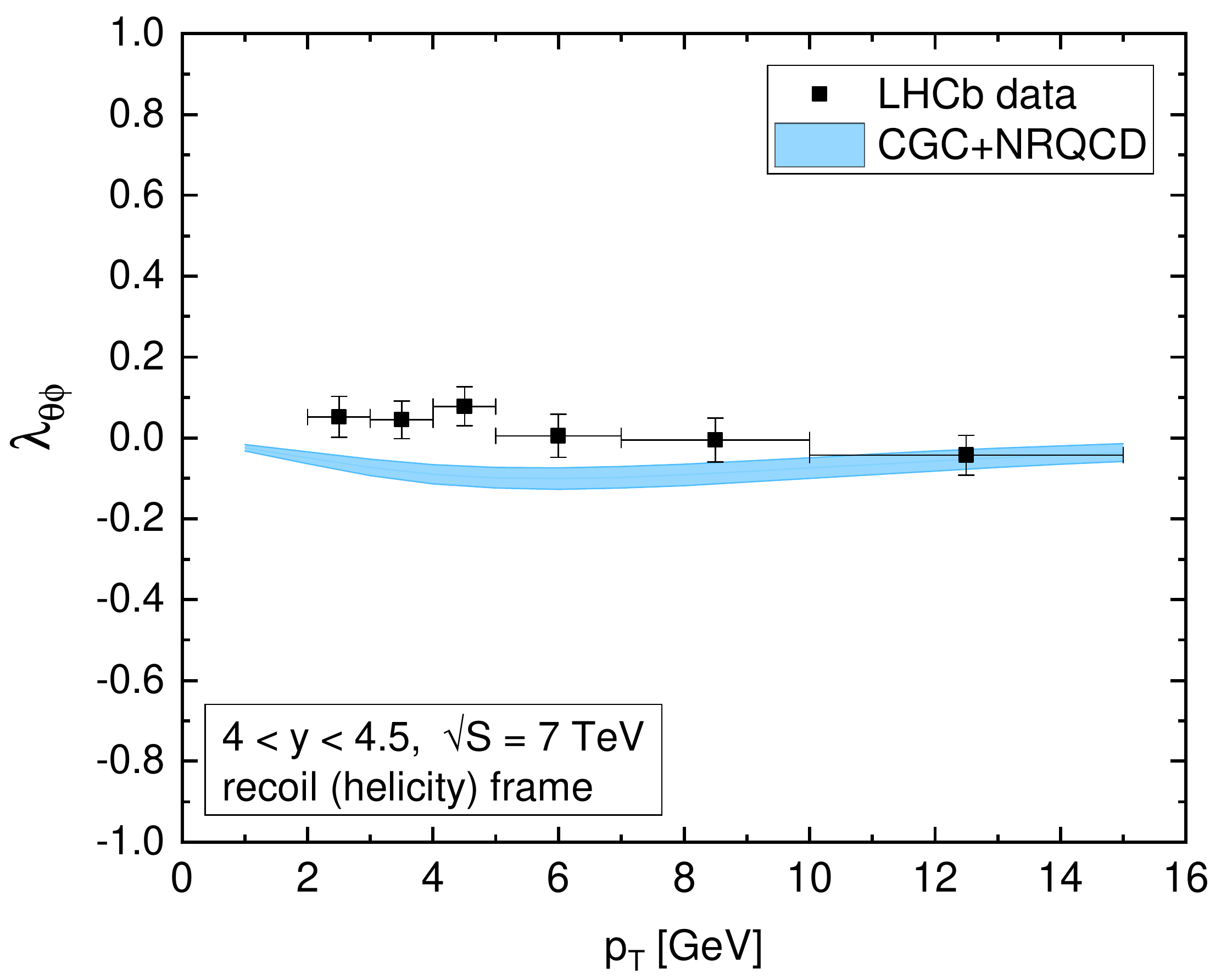}
\end{center}
\caption{The angular coefficients $\lamTheta$ (first row), $\lamPhi$ (second row) and $\lamThPh$ (third row) in the Collins-Soper frame (left column) and recoil frame (right column) as functions of $J/\psi$ transverse momentum $p_T$. Data are from the LHCb experiment at $7\tev$ \cite{Aaij:2013nlm}, for the highest rapidity window $4<y<4.5$.
}
\label{lambda_mv_LHCb_high_y}
\end{figure}

We can however compute the angular polarization variables as a function of rapidity and $p_T$ and compare these to data to check the quality of agreement with varying rapidity. In Figure \ref{lambda_mv_LHCb_y_dependence}, we show the rapidity dependence of all three coefficients plotted in both Collins--Soper and recoil frames. They are plotted for three transverse momentum values, $p_T=2.5, \ 4.5, \ 8.5 \gev$ represented by three bands on each plot (two of these are shifted by 0.5 and 1 unit for better visibility.) One observes that the CGC+NRQCD results are almost completely independent of rapidity in the range plotted. While this also appears to be the case for $\lamPhi$ and $\lamThPh$, the experimental values of $\lamTheta$ seem to have some rapidity dependence (within uncertainties) for $p_T=2.5$ and 4.5 GeV. In general, one may conclude that at higher rapidities our dilute-dense CGC+NRQCD predictions are closer to data as they should be. In Figure \ref{lambda_mv_LHCb_high_y}, we show LHCb data \cite{Aaij:2013nlm} collected for the highest rapidity window, $4<y<4.5$ and compare them with our predictions. One sees slightly better agreement for $\lamTheta$ than that seen in the wider rapidity window of $2.5<y<4$ plotted in Figure \ref{lambda_mv_LHCb_ALICE_7_8}. Though tempting, it would be premature however to conclude that this better agreement is primarily due to the dilute-dense approximation being better satisfied.

\begin{figure}
\begin{center}
\includegraphics[width=.48\textwidth]{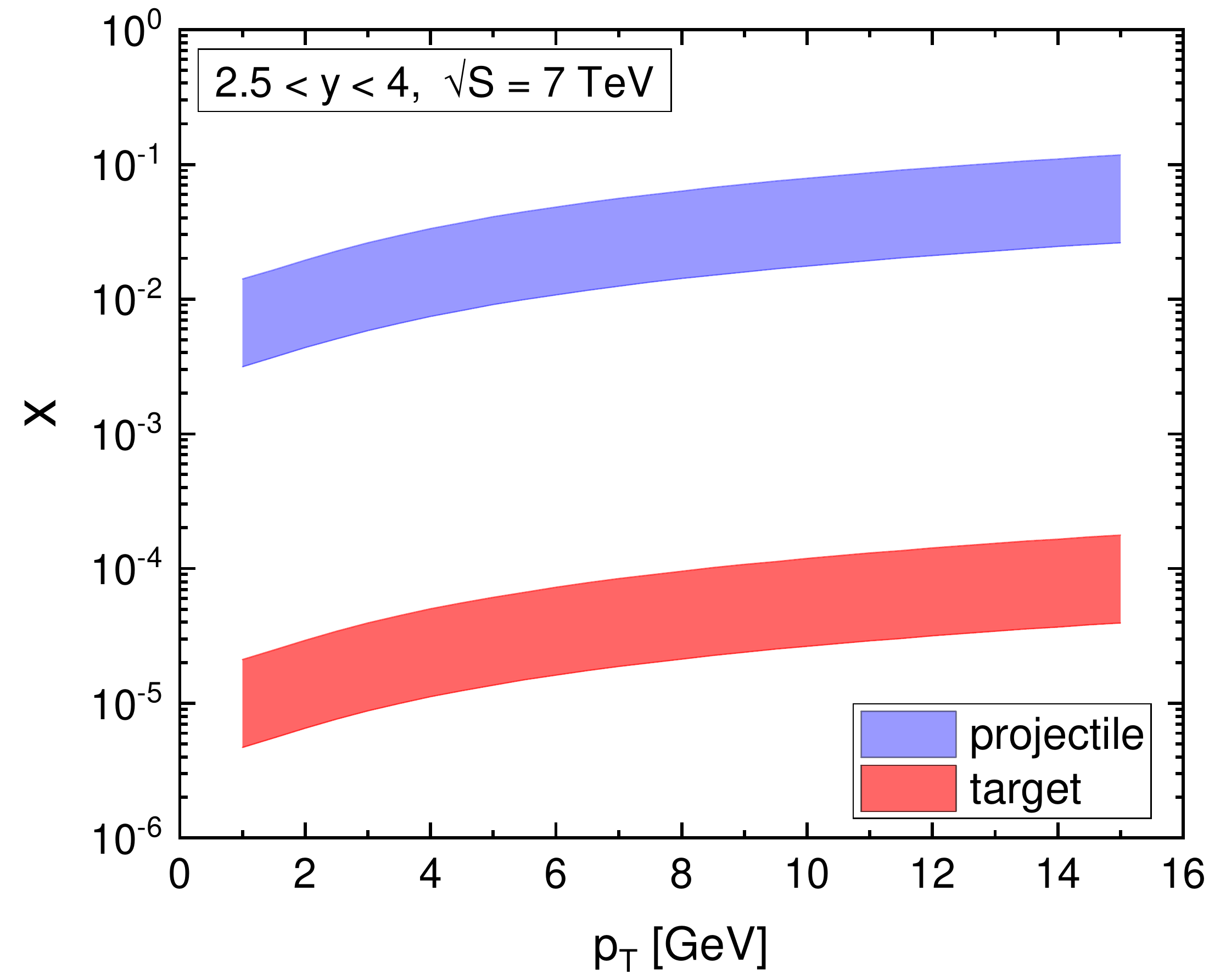}
\end{center}
\caption{Range of $x$ values probed in the $J/\psi$ production at forward rapidities, both for projectile and target protons. Bands are defined by the condition $2.5 < y < 4$.
}
\label{x_vs_pT}
\end{figure}

Note that our dilute-dense approximation in the CGC EFT assumes asymmetric treatment of the two colliding protons: the target is treated as a dense parton system for the resolved transverse momentum in the target. Similarly, the projectile is assumed to be dilute. This assumption is best satisfied for forward $J/\psi$ production at low $p_T$. In Fig.~\ref{x_vs_pT}, we show the kinematic range of $x$ values probed in the projectile and target. For the projectile, the $x$ values are quite large; however, as discussed in \cite{Ma:2014mri}, the unintegrated gluon distributions can be constrained by a smooth matching to the collinear pQCD gluon distribution. In the case of the target, the very small values of $x\sim 10^{-5}-10^{-4}$  motivate its treatment as a dense system. The kinematics of projectile and target suggest therefore that the application of the CGC dilute-dense formalism is appropriate. For $p_T\geq 8$ GeV, one is starting to probe $x$ values where replacing the unintegrated 
$k_T$ distribution with the gluon parton distribution will begin to receive large corrections. This situation will be exacerbated for $4<y<4.5$, where the hybrid formalism 
may be more appropriate; computations in such a framework matched to NRQCD are not available at present.

\subsection{Discussion}
\label{sec:Discussion}

The computation of the helicity SDCs in our paper are performed in the CGC weak coupling framework which, in principle, differs significantly from those in the NLO collinear factorization framework which has a  different kinematic window of applicability. However there is also a significant regime of overlap between the two approaches. As it was shown in \cite{Gelis:2003vh}, the dilute-dilute limit of the dilute-dense limit we have considered here is equivalent to the $k_T$ factorization formalism. Further, it was shown in  \cite{Gelis:2003vh} that for $k_T\rightarrow 0$ one can express the heavy-quark pair cross-section as a convolution of the LO collinear pQCD matrix element for $gg\rightarrow q\bar q$ scattering and the product of small $x$ gluon parton distributions. Thus in including the NLO BK evolution equation with 
running coupling~\cite{Balitsky:2008zza,Kovchegov:2006vj,Albacete:2007yr} in our 
computation of the cross-sections, we are including important pieces of the leading NLO, NNLO,$\cdots$ collinear pQCD contributions at small $x$, as previously also 
emphasized in \cite{Catani:1990xk,Catani:1990eg}. The matching between the two frameworks will fail when $p_T$ becomes sufficiently large that $p_T$ dependent contributions that are subleading in $x$ begin to play a role. 

\begin{figure}
\begin{center}
\includegraphics[width=.47\textwidth]{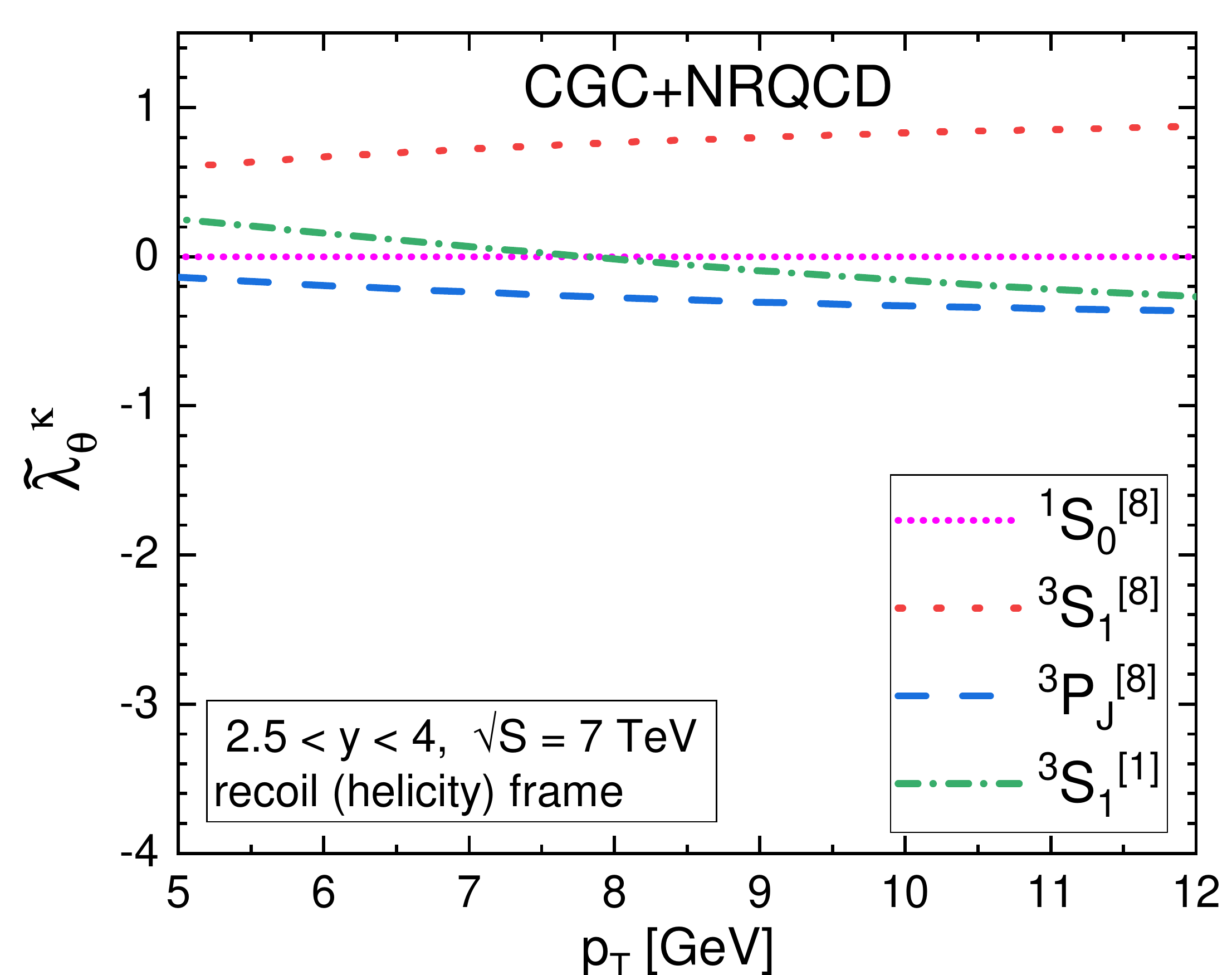}
\includegraphics[width=.47\textwidth]{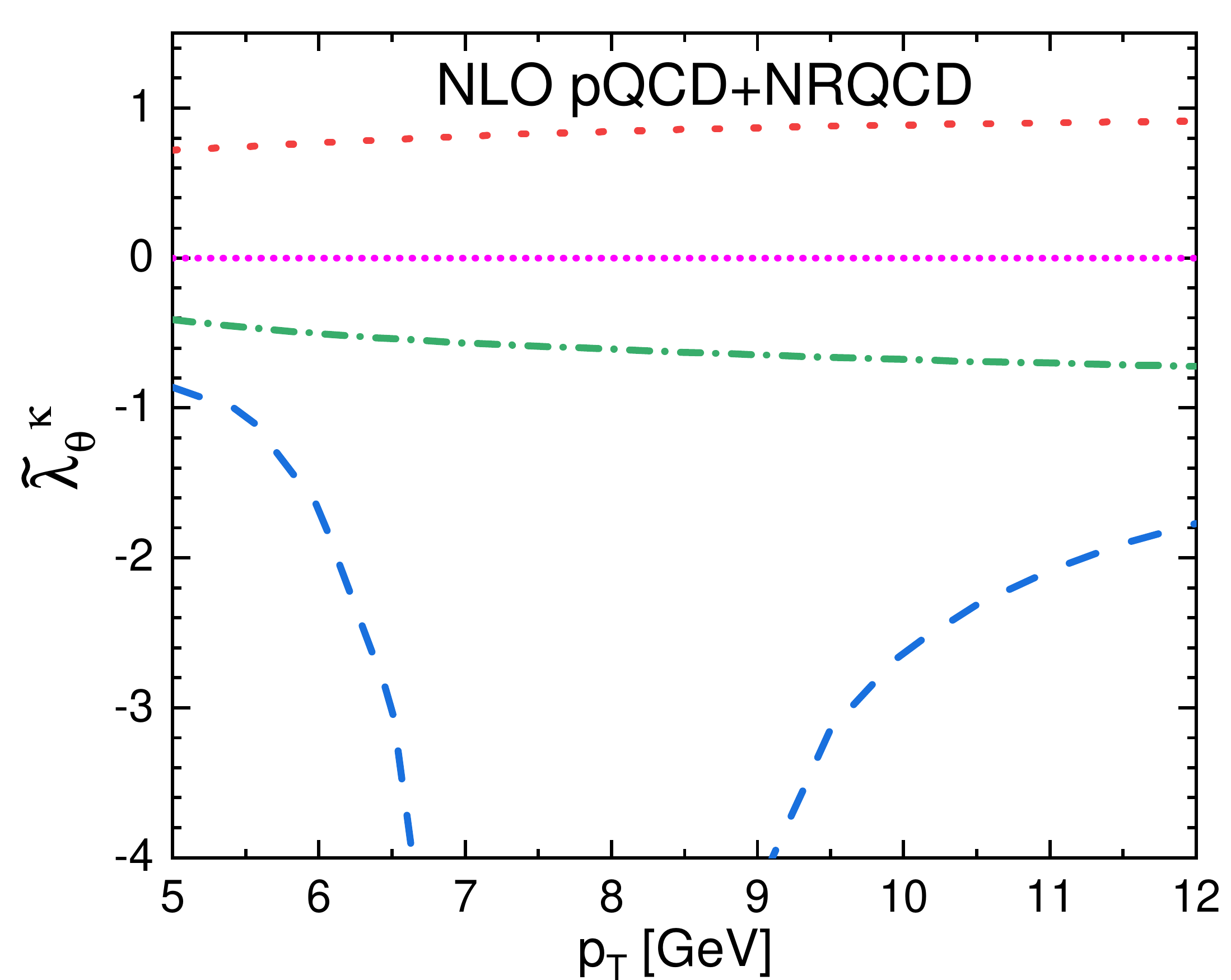}
\end{center}
\caption{The angular distribution coefficients $\tilde\lambda_\theta^\kappa$ in the recoil frame calculated for each channel separately using the CGC+NRQCD (left plot) and the NLO pQCD +NRQCD \cite{Shao:2014yta} (right plot) frameworks. For an extended discussion, see text.
}
\label{lambda_kappa_CGCvsNLO}
\end{figure}

It is at present not known analytically where in $p_T$ this mismatch occurs. This would require higher order computations in both frameworks than currently available. 
However one can see how good the matching is phenomenologically and where they begin to differ. Such a comparison was performed for $J/\Psi$ production in \cite{Ma:2014mri}; good agreement was obtained for $p_T \lesssim 10$GeV. For the case of $J/\Psi$ polarization, to illustrate the fact that our CGC computation includes important NLO collinear contributions, we will compare our result with the NLO pQCD+NRQCD calculation by Chao et al.~\cite{Chao:2012iv,Shao:2014yta}. We define the coefficient
\beq
\tilde\lambda_\theta^\kappa = \frac{d\hat\sigma^\kappa_{11}-d\hat\sigma^\kappa_{00}}{\left| d\hat\sigma^\kappa_{11}+d\hat\sigma^\kappa_{00} \right|},
\label{tilde_lambda}
\eeq
which is a polarization parameter $\lambda_\theta$ calculated for the given channel $\kappa$\footnote{We introduce absolute value in the denominator of (\ref{tilde_lambda}) because NLO pQCD $d\hat\sigma^{^3P_J^{[8]}}_{11}+d\hat\sigma^{^3P_J^{[8]}}_{00}$ decreases from being positive to negative as $p_T$ increases \cite{Chao:2012iv}. This sign change is the reason for $\tilde\lambda_\theta^{^3P_J^{[8]}}$'s divergence in Fig.~\ref{lambda_kappa_CGCvsNLO}}. Note that it does not depend on the values of the LDMEs. In Fig.~\ref{lambda_kappa_CGCvsNLO}, we show $\tilde\lambda_\theta^\kappa$'s in the recoil frame calculated using the two frameworks, CGC+NRQCD (left plot) and NLO pQCD+NRQCD (right plot). A characteristic property of the NLO pQCD+NRQCD computations is that the $^3S_1^{[1]}$ and $^3P_J^{[8]}$ channels give $\tilde\lambda_\theta^\kappa$ with a sign opposite to that for the $^3S_1^{[8]}$ channel; this cancellation leads to $\lambda_\theta\sim0$. As can be seen in 
Fig.~\ref{lambda_kappa_CGCvsNLO}, our CGC+NRQCD results also have a similar behavior confirming our expectation that the latter includes key physics of the NLO 
pQCD+NRQCD framework. We should emphasize that the divergence seen in Fig.~\ref{lambda_kappa_CGCvsNLO} for the $^3 P_J^{[8]}$ channel is not physical, since the quantity we 
are plotting is merely illustrative and is not what is measured in the $J/\Psi$ polarization studies.
In fact, we see that the CGC+pQCD computation is more stable than the NLO pQCD+NRQCD computation over the entire kinematic region shown. Note that the cancellations seen at NLO is not present~\cite{Chao:2012iv} for the LO pQCD+NRQCD computation.
To summarize, the most important difference relative to the NLO pQCD computations is that our framework includes higher twist gluon saturation contributions that become comparable to the leading twist contributions at small $x$ and low $p_T$. However at high $p_T$ our framework reduces to $k_T$ factorization which has a
significant regime of overlap in $p_T$ with higher order collinear pQCD computations.

The accuracy of the CGC+NRQCD computations should be significantly improved once NLO computations to the 
heavy quark-antiquark pair impact factor become available. There has been recent progress in this direction~\cite{Benic:2016uku,Benic:2018hvb,Roy:2018jxq} but much work remains to be done. An interesting question is the validity of the eikonal approximation that is assumed in the computation; these may potentially impact polarization observables more than unpolarized quantities. While there have been some efforts towards computing non-eikonal corrections in the CGC framework~\cite{Altinoluk:2014oxa,Altinoluk:2015gia,Altinoluk:2015xuy}, the application of these methods to quarkonium polarization is beyond the scope of this paper. 

We should note further that the results presented in this paper are obtained for direct $J/\psi$ production, whereas only prompt \cite{Aaij:2013nlm} or inclusive polarization data \cite{Abelev:2011md,Acharya:2018uww} are available. However our calculation is quite accurate because feeddown contributions to $J/\psi$ production from decays of higher charmonium states and $B$-hadrons are smaller at the low $p_T$'s that we consider \cite{LHCb:2012af,Aaij:2011jh}. Furthermore, an analysis within the collinear factorization pQCD+NRQCD framework suggests that these feeddown contributions do not change the result significantly \cite{Shao:2012fs}. Indeed, we have checked within the CGC framework itself that the feeddown corrections from $\chi_{cJ}$ states has a small impact on $J/\psi$ polarization parameters, and are within the estimated theoretical uncertainty band.

Finally, we note that the LDME set we are using was obtained from the fit to the data using SDCs calculated in the NLO collinear factorization framework \cite{Chao:2012iv}. An independent fit of LDMEs to the data using the SDCs computed in the CGC is possible; this program will however be most effective when the above mentioned NLO impact factors will become available. The good description of the data both for the yields \cite{Ma:2014mri} and for polarization variables suggests that such an exercise is very worthwhile and much needed. This is especially so because some of the LDME sets obtained in the NLO pQCD+NRQCD approach (for instance in \cite{Gong:2012ug} and \cite{Butenschoen:2011ks}) have negative values for some of the LDMEs. Using these LDMEs in our approach would lead to significant discrepancies between theory and data. A concern with using cross-sections at high $p_T$ is that small variations in $p_T$ can lead to large uncertainties in the LDMEs; a robust framework at lower values of $p_T$ can therefore greatly improve the precision with which they are extracted.

\section{Summary and outlook}

The CGC effective field theory has by now been used to successfully compute a large number of final states at collider energies. In this paper, we extended the CGC+NRQCD approach~\cite{Kang:2013hta} which was previously used to compute $J/\psi$ and $\psi(2S)$ yields in proton-proton collisions \cite{Ma:2014mri} to address the $J/\psi$ polarization puzzle. We have computed the three polarization coefficients $\lamTheta$, $\lamPhi$, $\lamThPh$ for proton-proton collisions in this approach and have obtained on the whole quite good results describing the $J/\psi$ polarization measured by the LHCb and ALICE experiments in both the Collins--Soper and recoil frames. 

In Section \ref{sec:Discussion}, we discussed the differences between our approach and that of collinearly factorized pQCD+NRQCD approaches, as well as some future refinements of the extant CGC+NRQCD computations. In particular, our results provide strong motivation to compute the NLO impact factor in the CGC+NRQCD EFT for quarkonium production in proton-proton and proton-nucleus collisions. The NLO impact factor results, when available, will allow for significant improvements in the accuracy of the extracted LDMEs; these at present differ considerably between different NLO pQCD+NRQCD analyses. 

Further, the increasing experimental precision of collider data opens up the possibility of computing the polarizations of the higher charmonium states in the CGC+NRQCD framework . In particular, the good description of data obtained for $\psi(2s)$ yields \cite{Ma:2014mri,Ma:2017rsu} suggests that this framework may also describe the polarization of this meson. Measurements of the $\chi_{cJ}$ states \cite{Aaij:2013dja} give access to the $ ^3P_J^{[1]}$ channel \cite{Ma:2010vd}. What more, the production of quarkonia containing $b$ quark pairs have been analyzed, in particular the polarization of $\Upsilon(ns)$ \cite{Aaij:2017egv}. The analysis of these states requires that 
we extend the CGC+NRQCD computations to resum large $\log (p_T/M)$ terms that appear in the perturbative computations \cite{Berger:2004cc,Sun:2012vc,Qiu:2013qka,Watanabe:2015yca}.

Finally, an analysis of $J/\psi$ polarization in high multiplicity events in proton-proton and proton-nucleus collisions is possible. First computations of the dependence of $J/\psi$ yields in proton-proton and proton-nucleus collisions were performed in \cite{Ma:2018bax} showing good agreement with data. The extension of this work studying the multiplicity dependence of the $J/\psi$ polarization coefficients is in progress and will be reported separately.

\paragraph{Acknowledgments}
We thank Michael Winn and Livio Bianchi for explaining to us details of the LHCb and ALICE polarization measurements. Y.-QM thanks Hao-Yu Liu, and TS thanks Renaud Boussarie, Krzysztof Golec-Biernat and Leszek Motyka, for useful discussions. Support of the Polish National Science Centre grants nos.\ DEC-2014/13/B/ST2/02486 is gratefully acknowledged. TS would also like to thank the Ministry of Science and Higher Education of Poland for support in the form of the Mobility Plus grant as well as Brookhaven National Laboratory for hospitality and support. RV is supported by the U.S. Department of Energy, Office of Science, Office of Nuclear Physics, under Contracts No. DE-SC0012704.

\appendix

\section{Explicit expressions for coefficients defining polarization frames}
\label{appendix_alpha_beta_coeff}
We provide here for completeness the coefficients defining $X$ and $Z$ axes in Eq.~(\ref{X_Z_def_by_A_B}) that were computed in \cite{Beneke:1998re} for the two frames employed. For the recoil (or helicity) frame these are,
\begin{eqnarray}
\alpha_z &=& -\frac{M}{\sqrt{(A\cdot p)^2-M^2 S}}, \qquad \qquad \beta_z=0\,,\nonumber \\
\alpha_x &=& \frac{A\cdot p\, B\cdot p}{\sqrt{S \,((A\cdot p)^2-M^2 S) ((A\cdot p)^2-(B\cdot p)^2-M^2 S)}}\,,\nonumber
\\
\beta_x &=& -\frac{\sqrt{(A\cdot p)^2-M^2 S}}{\sqrt{S\,((A\cdot p)^2-(B\cdot p)^2-M^2 S)}}\,.
\label{recoil_alpha_beta}
\end{eqnarray}
For the Collins--Soper frame they are given by,
\begin{eqnarray}
\alpha_z&=&-\frac{B\cdot p}{\sqrt{S\,((A\cdot p)^2-(B\cdot p)^2)}}\,, \qquad \nonumber
\beta_z=\frac{A\cdot p}{\sqrt{S \,((A\cdot p)^2-(B\cdot p)^2)}}\,, \\
\alpha_x &=& -\frac{M\,A\cdot p}
{\sqrt{((A\cdot p)^2-(B\cdot p)^2) \,((A\cdot p)^2-(B\cdot p)^2-M^2 S)}}\,,\nonumber
\\
\beta_x &=& \frac{M\,B\cdot p}
{\sqrt{((A\cdot p)^2-(B\cdot p)^2) \,((A\cdot p)^2-(B\cdot p)^2-M^2 S)}}\,,\label{CS_alpha_beta}
\end{eqnarray}
where $p$ is the quarkonium four--momentum, $S=(P_1+P_2)^2$ is hadron center of mass energy squared and $A, B$ are as defined in Eq.~(\ref{A_B_vec_def}).

\section{Computation of the CGC+NRQCD SDC}
\label{gamma_G_appendix}

In this section, we sketch the procedure of obtaining the functions $\Gamma^{\kappa}_{ij}$ and $ {\cal G}^\kappa_{ij}$ that contribute to the short distance coefficients of the spin density matrix, in a full analogy to that computed in \cite{Kang:2013hta,Ma:2014mri} for the CGC+NRQCD unpolarized quarkonium cross-sections.

\subsection{Amplitude}
We start by writing the amplitude obtained within the dilute-dense CGC formalism for $c\bar c$ pair production in a state $\kappa$ with fixed spin and orbital momentum projections $S_z$, $L_z$ and momentum $p$ (see Eq. (3.7) in \cite{Kang:2013hta}). This has the form,
\begin{align}\label{eq:amp}
\begin{split}
&M^{\kappa, (L_z,S_z)}(p)=\frac{g_s^2}{(2\pi)^4} \underset{\vka,\vk}{\int}
\frac{\rho_{p,a}(x_1,\vka)}{k_{1\perp}^2} \underset{\vx, \vy}\int
e^{i
\vk\cdot\vx} e^{i (\vp-\vk-\vka)\cdot\vy}\\
&\times \left\{ \text{Tr}\left[\mathcal{C}^{\kappa} V_F(\vx)t^a
V_F^\dagger(\vy)\right] \mathcal{F}^{\kappa,
(L_z,S_z)}_{Q\bar{Q}}\left(p,\vka,\vk\right) +
\text{Tr}\left[\mathcal{C}^{\kappa} t^b V_A^{b a}(\vx)\right]
\mathcal{F}^{\kappa, (L_z,S_z)}_g(p,\vka)\right\}\,,
\end{split}
\end{align}
where $V_F$ ($V_A$) is the Wilson line in the fundamental (adjoint) representation and $\rho_{p,a}(x_1,\vka)$ is the density of color sources in the proton.

The functions $\mathcal{F}^{\kappa,(L_z,S_z)}_{Q\bar{Q}/g}$ for the different spin helicity states are expressed as 
\bear
\mathcal{F}^{\ ^1S_0^{[8]},(0,0)}_{Q\bar{Q}}\left(p,\vka,\vk\right)&=& \left. \text{Tr}\left[\Pi^{00}(p,q)
T_{Q\bar{Q}}\left(p,q,\vka,\vk\right) \right]\right|_{q=0}, \\
\mathcal{F}^{\ ^3S_1 ,(0,S_z)}_{Q\bar{Q}/g}\left(p,\vka,\vk\right)&=&\epsilon^{*}_\mu(S_z) \left. \text{Tr}\left[\Pi^{\mu}(p,q) \,
T_{Q\bar{Q}/g}\left(p,q,\vka,\vk\right) \right]\right|_{q=0}, \\
\mathcal{F}^{\ ^3P_J^{[8]}, (L_z,S_z)}_{Q\bar{Q}}\left(p,\vka,\vk\right)&=&
\epsilon^{*}_\mu(S_z) \epsilon^{*}_\beta(L_z) \left. \frac{\partial}{\partial q^{\beta}}
\text{Tr}\left[\Pi^{\mu}
T_{Q\bar{Q}}\left(p,q,\vka,\vk\right) \right]\right|_{q=0}.
\enar
These functions represent the action of the covariant spin projectors \cite{Kuhn:1979bb,Guberina:1980dc}
\bear
\Pi^{00}&=&\frac{1}{\sqrt{8m^3}}\left(\frac{\slashed{p}}{2}-\slashed{q}-m\right)
\gamma^5\left(\frac{\slashed{p}}{2}+\slashed{q}+m\right)\,, \\
\Pi^{\mu}&=&\frac{1}{\sqrt{8m^3}}\left(\frac{\slashed{p}}{2}-\slashed{q}-m\right)
\gamma^\mu \left(\frac{\slashed{p}}{2}+\slashed{q}+m\right)\,,
\enar
on the amplitudes $T_{Q\bar{Q}}\left(p,q,\vka,\vk\right)$, $T_g\left(p,q,\vka,\vk\right)$ for the process $g(k_1) \rightarrow Q\left(\frac{p}{2}+q\right) \bar{Q}\left(\frac{p}{2}-q\right)$ with Wilson lines attached to quarks or gluons respectively \cite{Blaizot:2004wv}. Explicit expressions for the latter can be found in Eq.~(2.9) of \cite{Kang:2013hta}. Finally, the quarkonium polarization vectors $\epsilon^\mu$ can be expressed in terms of the unit axis vectors $X,Y,Z$ in the $J/\psi$'s rest frame defined in section \ref{ang_dist_section} and are given by 
\begin{equation}
\epsilon^\mu(0)=Z^\mu, \qquad
\epsilon^\mu(\pm 1)=\frac{1}{\sqrt{2}}(\mp X^\mu-i Y^\mu)\,. 
\end{equation}

\subsection{Functions $\Gamma^{\kappa}_{ij}$ and $ {\cal G}^\kappa_{ij}$}

The functions $\Gamma^{\kappa}_{ij}$ and $ {\cal G}^\kappa_{ij}$ appearing in Eqs.~(\ref{eq:dsktCO}) and (\ref{eq:dsktCS}) are obtained by taking the modulus squared of the amplitude in Eq.~(\ref{eq:amp}). The averaging over the color source densities in the projectile and target results in the unintegrated gluon densities and dipole amplitudes respectively, as described in \cite{Kang:2013hta,Ma:2014mri}. The rest of the expression, containing all the information on the helicities then condensed into 
\bear
\Gamma^{\, ^1S_0^{[8]}}_{ij}\left(p,\vka,\vk\right)&=& \frac{1}{3} \delta_{ij} \left| \left. \text{Tr}\left[\Pi^{00}
T_{Q\bar{Q}}\right]\right|_{q=0} \right|^2\, \\
\Gamma^{^3S_1^{[8]}}_{ij}\left(x_1,x_2, p,\vka,\vk\right)&=&\epsilon^{*}_\mu(i)\epsilon_\nu(j)\frac{1}{3} \left. \text{Tr}\left[\Pi^{\mu} \,
(T_{Q\bar{Q}}+T_g)\right]\right|_{q=0} \left( \left. \text{Tr}\left[\Pi^{\nu} \,
(T_{Q\bar{Q}}+T_g)\right]\right|_{q=0} \right)^*,\\
\Gamma^{^3P_J^{[8]}}_{ij}\left(x_1,x_2, p,\vka,\vk\right)&=& \frac{1}{9} \left( \sum_{L_z=-1}^1
\epsilon^{*}_\beta(L_z) \epsilon_{\alpha}(L_z) \right)  \nonumber \\
&\times&\epsilon^{*}_\mu(i)\epsilon_\nu(j) \left. \frac{\partial}{\partial q^{\beta}}
\text{Tr}\left[\Pi^{\mu}T_{Q\bar{Q}}\right]\right|_{q=0}
\left( \left. \frac{\partial}{\partial q^{\alpha}}
\text{Tr}\left[\Pi^{\nu}T_{Q\bar{Q}}\right]\right|_{q=0} \right)^*\,, \\
{\cal G}^{^3S_1^{[1]}}_{ij}\left(x_1,x_2, p,\vka,\vk,\vkp \right)&=& \epsilon^{*}_\mu(i)\epsilon_\nu(j)
\frac{1}{6} \left. \text{Tr}\left\{ \Pi^{\mu} \, \left[T_{Q\bar{Q}}\left(p,q,\vka,\vk\right)-T_{Q\bar{Q}}\left(p,q,\vka,\vkp\right) \right] \right\}\right|_{q=0} \nonumber \\
&\times& \left( \left. \text{Tr}\left\{ \Pi^{\nu} \, \left[T_{Q\bar{Q}}\left(p,q,\vka,\vk\right)-T_{Q\bar{Q}}\left(p,q,\vka,\vkp\right) \right] \right\}\right|_{q=0}\right)^*\,.
\enar
One needs to evaluate the Dirac traces of the form $\left.\text{Tr}\left[\Pi^{00} \,T_{Q\bar{Q}}\right]\right|_{q=0}$, $\left.\text{Tr}\left[\Pi^{\mu} \,T_{Q\bar{Q}}\right]\right|_{q=0}$, $\left.\text{Tr}\left[\Pi^{\mu} \,T_{g}\right]\right|_{q=0}$ and $ \left. \frac{\partial}{\partial q^{\alpha}} \text{Tr}\left[\Pi^{\nu}T_{Q\bar{Q}}\right]\right|_{q=0}$ and then contract them with polarization vectors $\epsilon^\mu$. We do not show the explicit results for these traces because the expressions are lengthy but will provide them upon request.

\section{Contributions from different states to $\lambda$ coefficients} 

To obtain further insight into how different channels contribute to the polarization coefficients $\lambda$, we define the variables $\lambda^\kappa$; these quantities have only contributions from given channel $\kappa$ in the numerators while the denominator have contributions from all channels:
\beq
\lamTheta^\kappa=\frac{d\sigma^\kappa_{11}-d\sigma^\kappa_{00}}{d\sigma_{11}+d\sigma_{00}}\,, \hspace{1cm}
\lamPhi^\kappa=\frac{d\sigma^\kappa_{1,-1}}{d\sigma_{11}+d\sigma_{00}}\,, \hspace{1cm}
\lamThPh^\kappa=\frac{\sqrt{2}\; \rm{Re}(d\sigma^\kappa_{10}) }{d\sigma_{11}+d\sigma_{00}}\,.
\label{lam_kappa_Defin}
\eeq
Note that we have denoted here $d\sigma_{ij}=\sum_\kappa d\sigma^\kappa_{ij}$ to ensure that the denominators in Eq.~(\ref{lam_kappa_Defin}) are the same as those in Eq.~(\ref{lam_Defin}).
These definitions then clearly satisfy
\beq
\sum_\kappa \lamTheta^\kappa = \lamTheta\,, \hspace{1cm} 
\sum_\kappa \lamPhi^\kappa = \lamPhi\,, \hspace{1cm} 
\sum_\kappa \lamThPh^\kappa = \lamThPh\,.
\eeq

\begin{figure}
\begin{center}
\includegraphics[width=.47\textwidth]{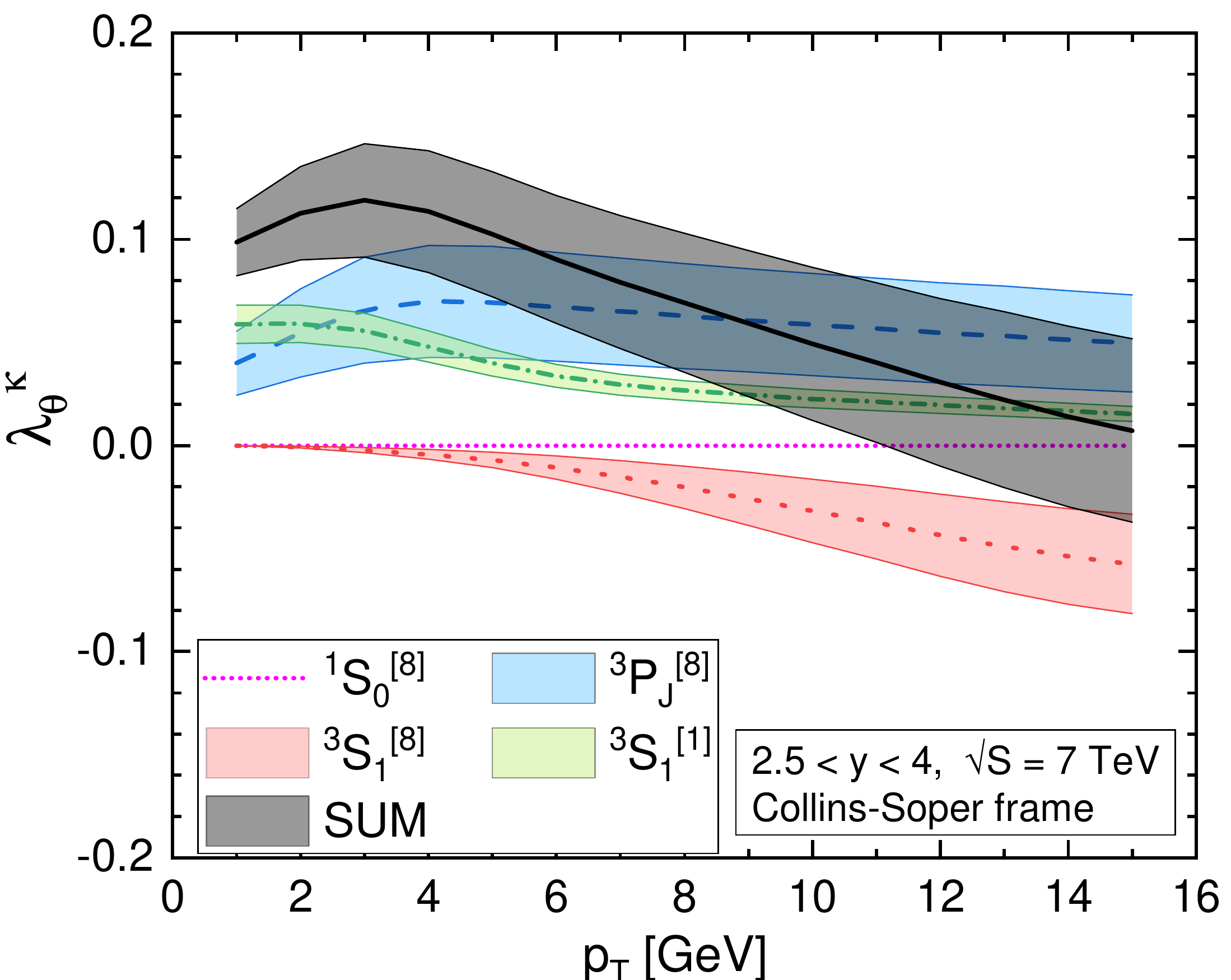}
\vspace{0.2cm}
\includegraphics[width=.47\textwidth]{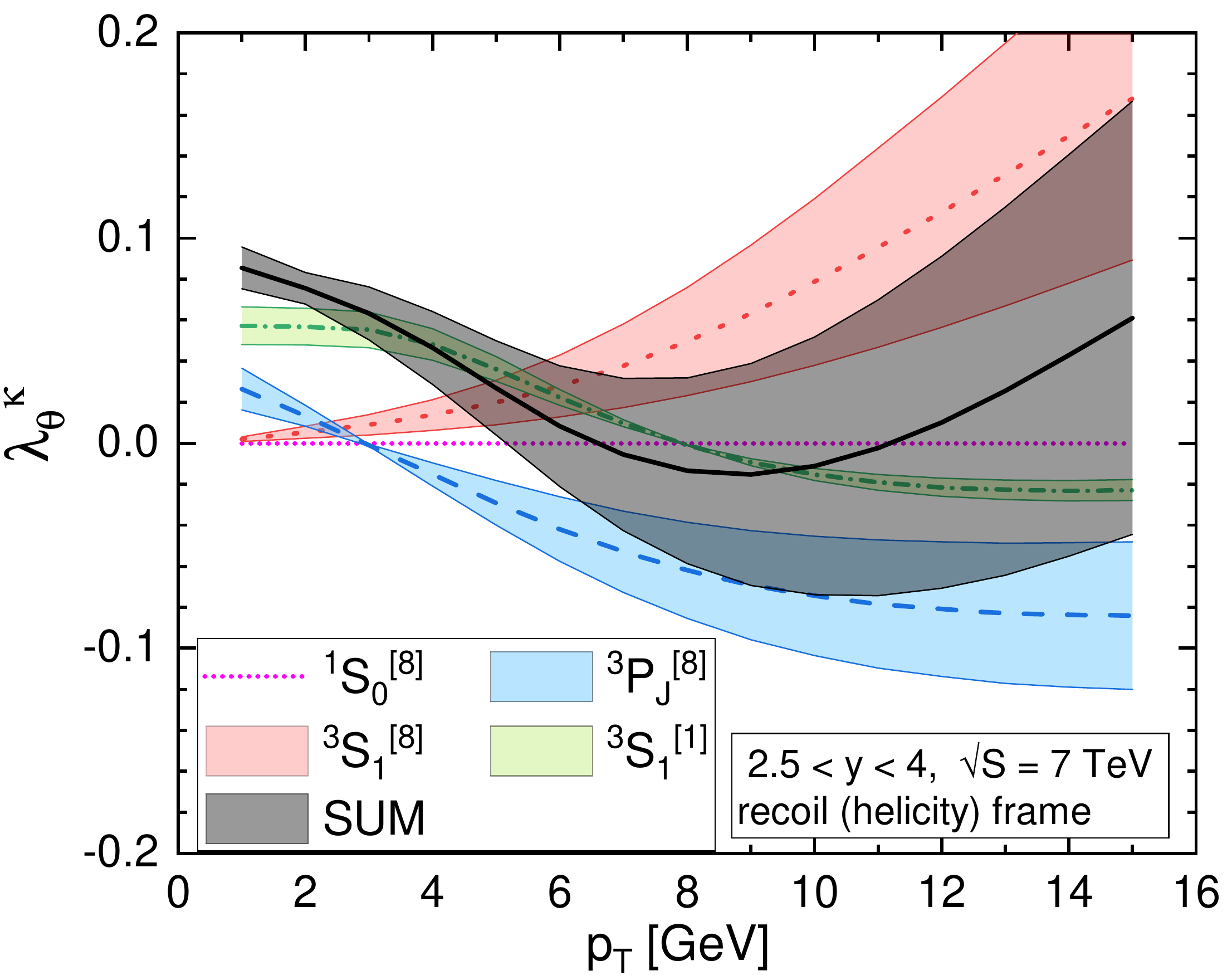}
\vspace{0.2cm}
\includegraphics[width=.47\textwidth]{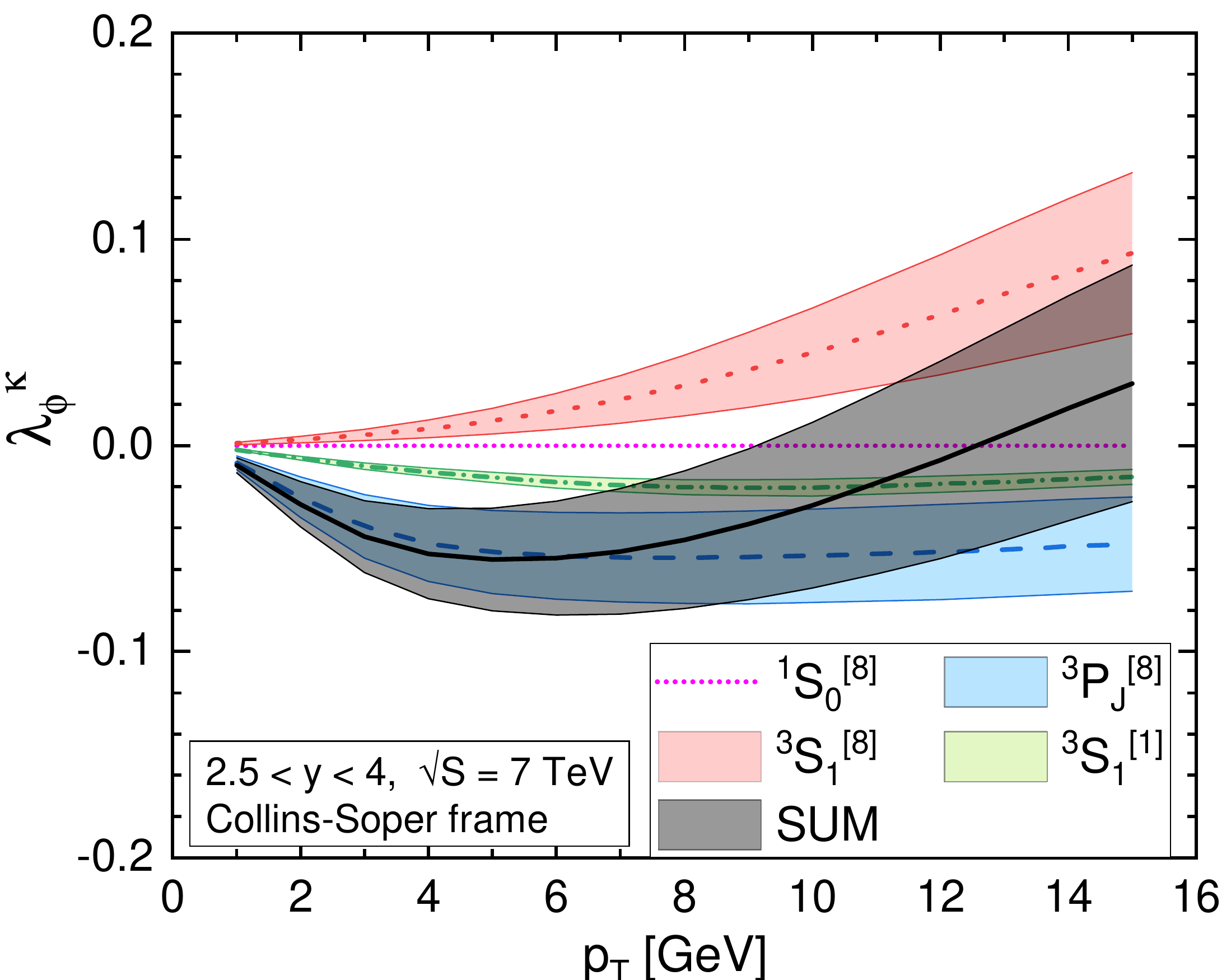}
\includegraphics[width=.47\textwidth]{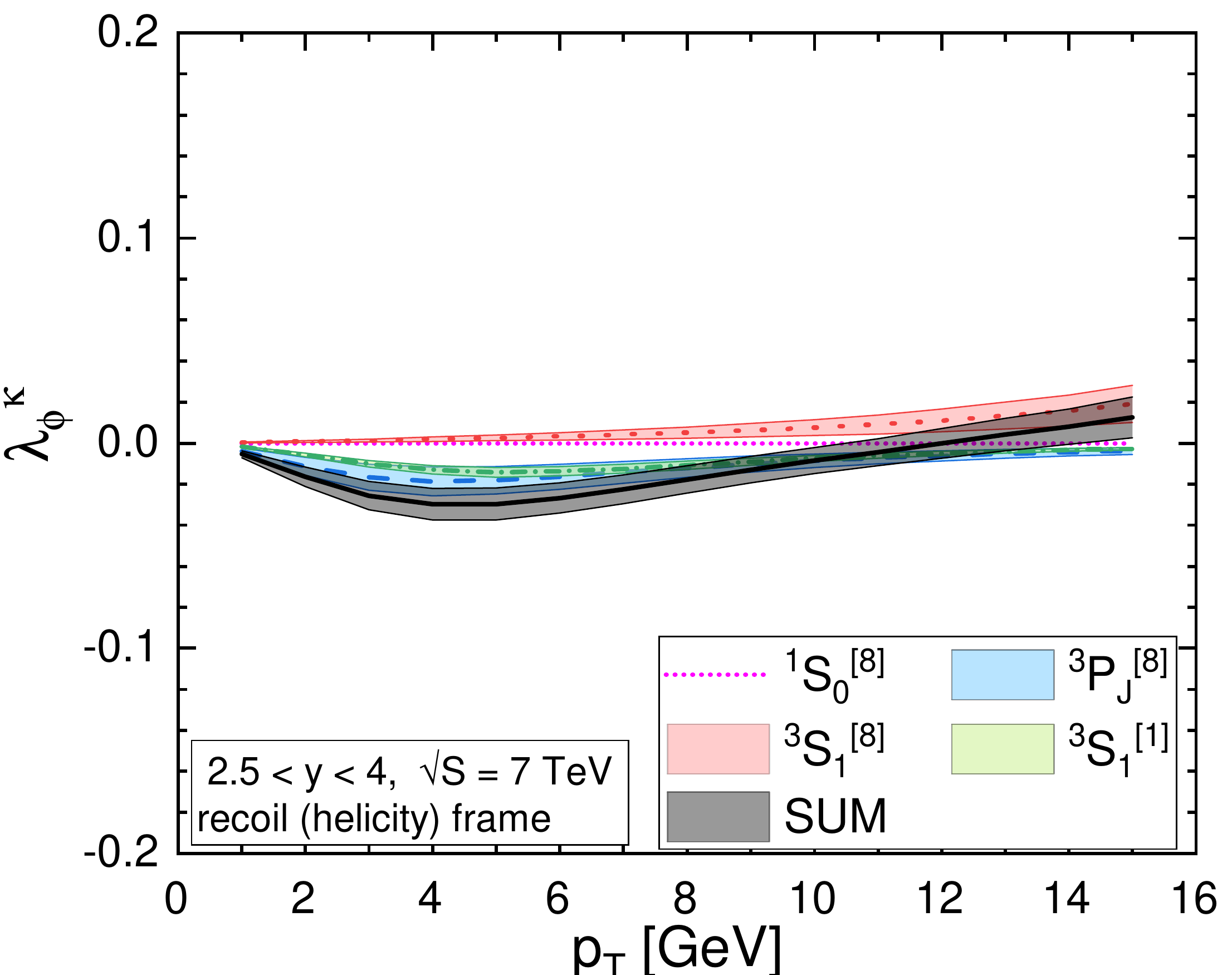}
\includegraphics[width=.47\textwidth]{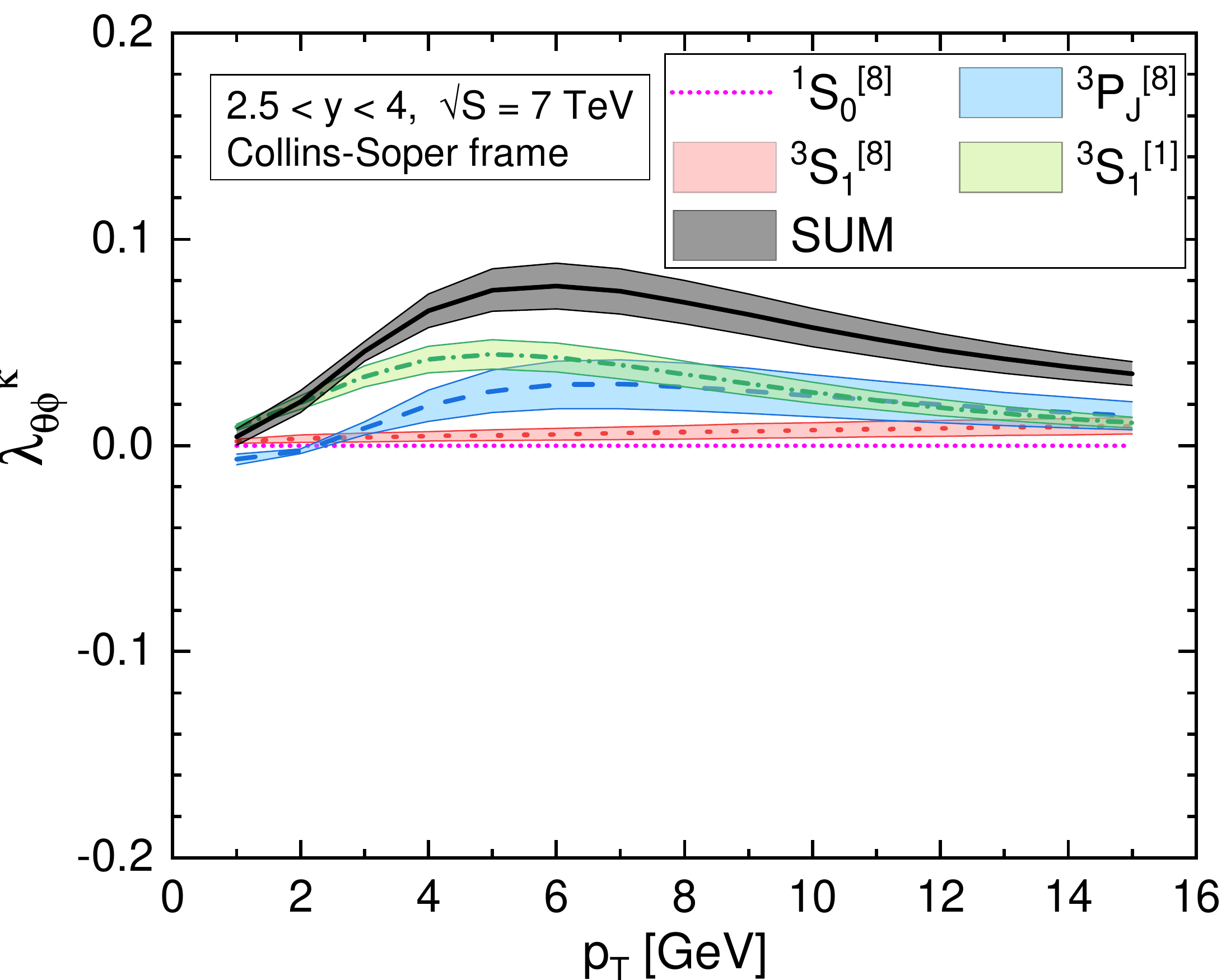}
\includegraphics[width=.47\textwidth]{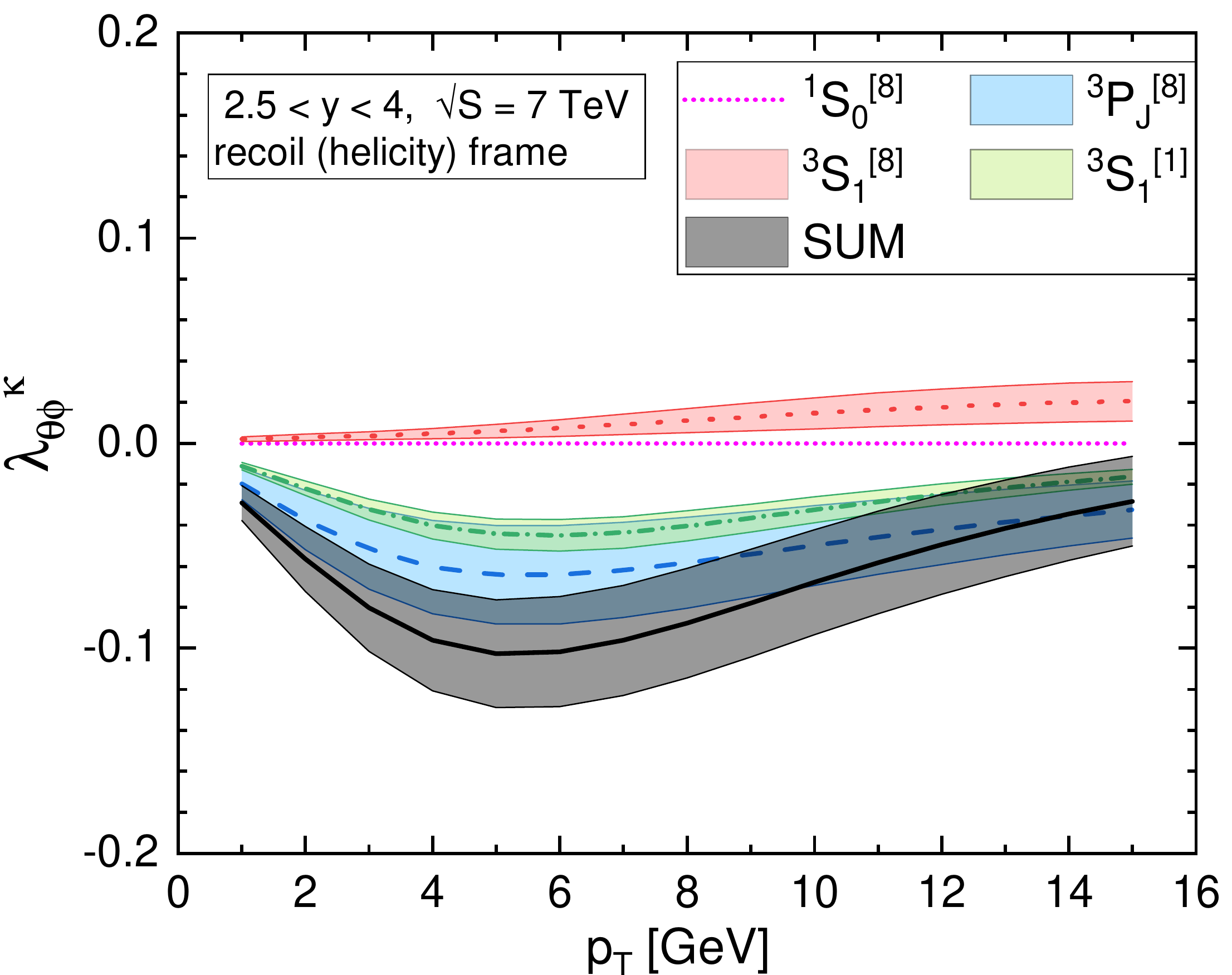}
\end{center}
\caption{Parameters $\lamTheta^\kappa$ (first row), $\lamPhi^\kappa$ (second row) and $\lamThPh^\kappa$ (third row) defined in Eq.~(\ref{lam_kappa_Defin}) plotted for the Collins-Soper frame (left column) and recoil frame (right column). Different colors of bands represents different channels. "SUM" (denoted as black band) represents sum of all channels, so is equal to $\lamTheta$, $\lamPhi$, $\lamThPh$ from Figure~\ref{lambda_mv_LHCb_ALICE_7_8} (note the difference in scales between these plots and those from Figure~\ref{lambda_mv_LHCb_ALICE_7_8}).
}
\label{lambda_kappa_plot}
\end{figure}

Our results for $\lambda^\kappa$ are shown in Figure \ref{lambda_kappa_plot}. We first observe that for $p_T \rightarrow 0$ all $\lamPhi^\kappa$ and $\lamThPh^\kappa$ vanish. This is as anticipated because in this limit azimuthal symmetry is restored and no $\phi$--dependence in Eq.~(\ref{ang_distribution_in_Jpsi_frame}) is possible. Note that the $^1S_0^{[8]}$ state does not contribute to the $\lambda$'s as we discussed in section \ref{den_matrx_sect}. At low $p_T$, the dominant contributions to the polarization coefficients come from the $^3S_1^{[1]}$ and $^3P_J^{[8]}$ channels which add up with the same sign in most cases. At higher $p_T$, the $^3S_1^{[8]}$ state becomes important. In most cases, it contributes with the opposite sign relative to the $^3S_1^{[1]}$ and $^3P_J^{[8]}$ states. This is especially striking for $\lambda_\theta$, where the $^3S_1^{[8]}$ state contributes with a transverse polarization while $^3P_J^{[8]}$ state is longitudinally polarization, with the summation of the two giving a nearly unpolarized result. As noted in the introduction to this paper, this phenomenon is very similar to that observed in the NLO collinear factorization framework. So albeit the CGC computation is computed at LO in the impact factor, our 
result suggests that it may contain key features of the dynamics of the NLO collinear factorization framework. This was also seen in the close matching of the yields of the unpolarized cross-section observed in \cite{Ma:2014mri}.

\bibliographystyle{JHEP}

\bibliography{biblio_Jpsi_CGC}

\end{document}